%% file: 0-Main.tex
\pdfoutput=1
\documentclass[acmsmall]{acmart}
\setcopyright{rightsretained}
\acmDOI{10.1145/3632923}
\acmYear{2024}
\copyrightyear{2024}
\acmSubmissionID{popl24main-p544-p}
\acmJournal{PACMPL}
\acmVolume{8}
\acmNumber{POPL}
\acmArticle{81}
\acmMonth{1}
\received{2023-07-11}
\received[accepted]{2023-11-07}
\usepackage{macros}

\let\svthefootnote\thefootnote
\newcommand\freefootnote[1]{%
  \let\thefootnote\relax%
  \footnotetext{#1}%
  \let\thefootnote\svthefootnote%
}

\begin{document}

\title[SimuQ: A Framework for Programming Quantum Hamiltonian Simulation with Analog Compilation]{SimuQ: A Framework for Programming Quantum Hamiltonian Simulation with Analog Compilation \\ (Extended Version)}



\author[Y. Peng]{Yuxiang Peng}
\affiliation{
  \institution{University of Maryland}
  \city{College Park}
  \state{Maryland}
  \country{United States}
}
\orcid{0000-0003-0592-7131}
\email{ypeng15@umd.edu}

\author[J. Young]{Jacob Young}
\affiliation{
  \institution{University of Maryland}
  \city{College Park}
  \state{Maryland}
  \country{United States}
}
\orcid{0009-0002-9133-1560}
\email{jyoung25@umd.edu}

\author[P. Liu]{Pengyu Liu}
\affiliation{
  \institution{Carnegie Mellon University}
  \city{Pittsburgh}
  \state{Pennsylvania}
  \country{United States}
}
\orcid{0000-0003-1302-2391}
\authornote{Pengyu Liu started participating in this work when he was an undergraduate at Tsinghua University, Beijing, China.}
\email{pengyul@andrew.cmu.edu}

\author[X. Wu]{Xiaodi Wu}
\affiliation{
  \institution{University of Maryland}
  \city{College Park}
  \state{Maryland}
  \country{United States}
}
\orcid{0000-0001-8877-9802}
\email{xwu@cs.umd.edu}

\begin{abstract}
Quantum Hamiltonian simulation, which simulates the evolution of quantum systems and probes quantum phenomena, is one of the most promising applications of quantum computing. 
Recent experimental results suggest that Hamiltonian-oriented analog quantum simulation would be advantageous over circuit-oriented digital quantum simulation in the Noisy Intermediate-Scale Quantum (NISQ) machine era. 
However, programming analog quantum simulators is much more challenging due to the lack of a unified interface between hardware and software. 
In this paper, we design and implement SimuQ, the first framework for quantum Hamiltonian simulation that supports Hamiltonian programming and pulse-level compilation to heterogeneous analog quantum simulators. 
Specifically, in SimuQ, front-end users specify the target quantum system with Hamiltonian Modeling Language, and the Hamiltonian-level programmability of analog quantum simulators is specified through a new abstraction called the abstract analog instruction set (AAIS) and programmed in AAIS Specification Language by hardware providers. 
Through a solver-based compilation, SimuQ generates executable pulse schedules for real devices to simulate the evolution of desired quantum systems, which is demonstrated on superconducting (IBM), neutral-atom (QuEra), and trapped-ion (IonQ) quantum devices.
Moreover, we demonstrate the advantages of exposing the Hamiltonian-level programmability of devices with native operations or interaction-based gates and establish a small benchmark of quantum simulation to evaluate SimuQ's compiler with the above analog quantum simulators.
\end{abstract}

\newif\iftechrep\techreptrue 



\maketitle

\input{1-Introduction}

\input{2-Example}

\input{3-DSL}

\input{4-Compiler}

\input{5-CaseStudy}

\input{6-Conclusion}

\begin{acks}
We thank the anonymous reviewers for their constructive feedback and thank Robert Rand, Kesha Hietala, Jens Palsberg, Frederic Chong, Cedric Lin, Peter Komar, Cody Wang, Jean-Christophe Jaskula,  Murphy Niu, Lei Fan, and Yufei Ding for their helpful discussions. Y.P., J.Y., and X.W. were partially funded by the U.S. Department of Energy, Office of Science, Office of Advanced Scientific Computing Research, Quantum Testbed Pathfinder Program under Award Number DE-SC0019040, Air Force Office of Scientific Research under award number FA9550-21-1-0209, the U.S. National Science Foundation grant CCF-1942837 (CAREER), and a Sloan research fellowship. This research used resources of the Oak Ridge Leadership Computing Facility, which is a DOE Office of Science User Facility supported under Contract DE-AC05-00OR22725.
\end{acks}

\section*{Data Availability Statement}
Our code is available at \url{https://github.com/PicksPeng/SimuQ}. \\
A project website of SimuQ is available at \url{https://pickspeng.github.io/SimuQ/}.

\bibliography{refs}

\newpage
\appendix
\input{9-Appendix.tex}

\end{document}

%% file: 1-Introduction.tex
\section{Introduction}

\subsection{Background and Motivation} 

Developing appropriate abstraction is a critical step in designing programming languages that help bridge the domain users and the potentially complicated computing devices. Abstraction is a fundamental factor in the productivity of the underlying programming language. 
Prominent early examples of such include, e.g., FORTRAN~\cite{fortran-history} and SIMULA~\cite{simula-history}, both of which provide high-level abstractions for modeling desirable operations for domain applications and have been proven enormous successes in history. 

Conventionally, abstractions for quantum computing adopt (qubit-level) quantum circuits to describe procedures, a mathematically simple approach that works well as a mental tool for the theoretical study of quantum information and algorithms \cite{nielsen2002quantum, childs2017lecture}. 
As a result, many quantum programming languages \cite
{Green2013, abhari2012scaffold, hietala2021verified, paykin2017qwire} have adopted quantum circuits as the only abstraction. Many quantum applications are implemented using these programming languages to generate quantum circuits, although only a few can be demonstrated on existing quantum devices.

\vspace{3mm}
\emph{Quantum Hamiltonian simulation} (also called quantum simulation\footnote{In certain contexts, \emph{quantum simulation} and \emph{quantum simulators} refer to the classical simulation of quantum circuits and the corresponding classical software tools, respectively. Yet throughout this paper, quantum simulation represents the task of simulating a quantum Hamiltonian system, and quantum simulators represent controllable quantum devices that are capable of simulating other quantum systems. }) is arguably one of the most promising quantum applications. The evolution of a quantum system, starting from a quantum state represented by a high-dimensional complex vector $\ket{\psi(0)}$, obeys the \emph{Schr\"odinger} equation:
\begin{align}
    \label{eq:sch}
    \frac{\d}{\d t}\ket{\psi(t)}=-iH(t)\ket{\psi(t)},
\end{align}
where $H(t)$ is generally a time-dependent Hermitian matrix, also known as the \emph{Hamiltonian} governing the system. Probing quantum phenomena from solutions of the \emph{Schr\"odinger} equation is a promising approach to tackle many open problems in various domains, including quantum chemistry, high-energy physics, and condensed matter physics \cite{cao2019quantum, nachman2021quantum, hofstetter2018quantum}. However, for an $n$ qubit system, the dimension of both $H(t)$ and $\ket{\psi(t)}$ could be $2^n$, which makes its classical simulation exponentially difficult in general. Though mature software developments for classical simulation of quantum systems using methods like quantum Monte Carlo \cite{foulkes2001quantum} and density-matrix renormalization groups \cite{schollwock2005density, schollwock2011density} succeed for restricted cases, many intermediate-size ($\sim$100 sites) quantum systems of significance are still out of reach for classical computers.

\begin{figure}
    \centering    
    \includegraphics[width=0.9\linewidth, trim=4cm 4.5cm 3.5cm 6cm, clip]{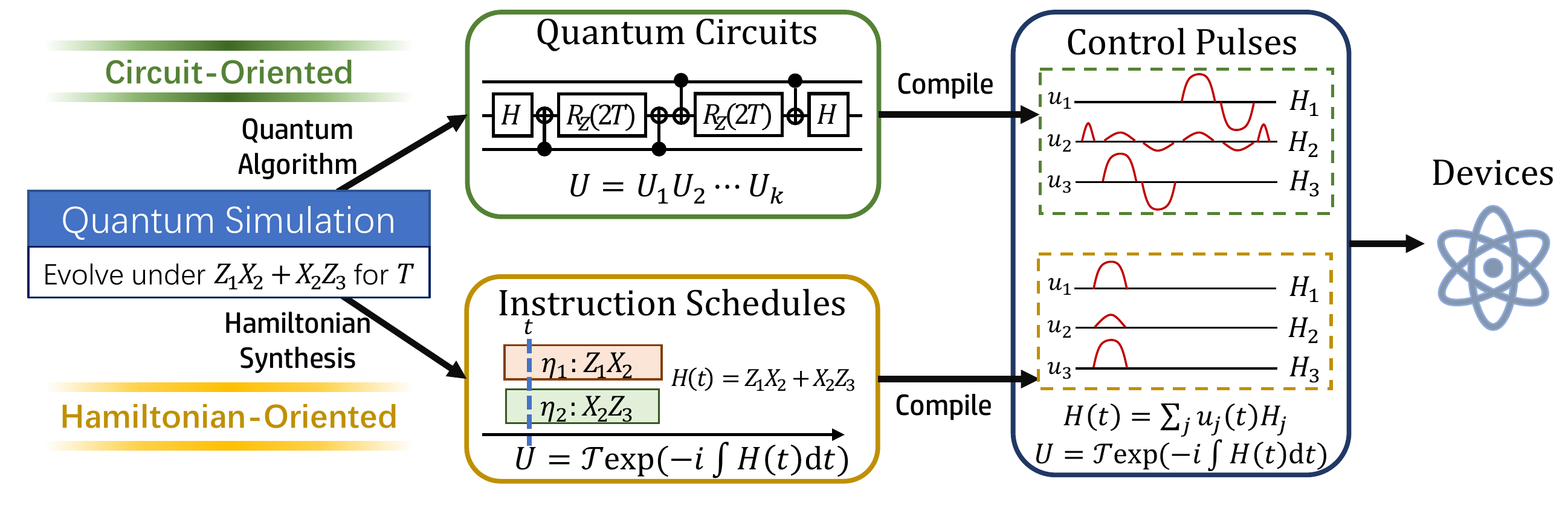}
    \vspace{-1.5em}
    \caption{The circuit-oriented and Hamiltonian-oriented schemes for compiling quantum Hamiltonian simulation on quantum devices. Here $\mathcal{T}\mathrm{exp}(-i\int H(t)\mathrm{d}t)$ is a solution to a Schr\"odinger equation governed by $H(t)$. }
    \vspace{-1em}
    \label{fig:cH-comp}
\end{figure}

To address this issue, in his famous 1981 lecture, \citet{feynman1982simulating} suggested employing a precisely controlled quantum system to simulate a target quantum system to avoid exponential complexity. Modern quantum technologies foster a variety of platforms to advance the realization of Feynman's proposal, for example, photonic systems \cite{o2009photonic}, superconducting circuits \cite{wendin2017quantum}, semiconductor nanocrystals \cite{kloeffel2013prospects}, neutral atom arrays \cite{saffman2016quantum}, and trapped-ion arrays \cite{bruzewicz2019trapped}. Most of them are described by a Hamiltonian with continuous-time parameters characterizing the signals sent through controllable physics instruments like microwaves or magnetic fields. They are called \emph{analog quantum simulators}. Only a few devices support a specific set of system evolutions with sophisticated pulse engineering, abstracted as a set of universal quantum gates \cite{kitaev1997quantum}, hence called \emph{digital quantum computers}. They include IBM's superconducting devices \cite{cross2018ibm} and IonQ's trapped-ion devices \cite{debnath2016demonstration}. However, inherent noises on near-term digital quantum computers induce detrimental errors causing short coherence time (i.e., quantum states do not deteriorate to classical states within it) and preventing demonstrating large quantum applications with provable speedup. The solution through fault-tolerant quantum computing \cite{gottesman2010introduction} requires significantly lower gate implementation errors and better device connectivity, impractical in the NISQ era \cite{preskill2018quantum}.

In the past decades, efficient quantum algorithms for quantum Hamiltonian simulation have been proposed \cite{lloyd1996universal, childs2010relationship, childs2012hamiltonian, low2017optimal, lauvergnat2007simple}. Implementing them follows a circuit-oriented scheme, where the quantum algorithms are designed and programmed as quantum circuits consisting of quantum gates abstracting the evolution of a fraction of sites in the quantum system. Then a quantum circuit compiler rewrites the circuits using a small set of quantum gates and translates each gate to pulses for specific devices. However, both programming and deploying algorithms in this scheme are highly non-trivial. In a seminal project, \citet{pnas-simulation} spent nearly two years programming a few major quantum simulation algorithms in Quipper~\cite{Green2013} due to tedious implementation details at the circuit level. Meanwhile, \cite{pnas-simulation} shows that implementing algorithms via quantum circuits, even for a simple quantum system of medium sizes (around 100 qubits), requires an astronomical number of gates (around $10^{10}$ before fault-tolerant encoding). Such circuits are far out of the reach of near-term quantum devices, which at most support a few thousand physical gates. The redundancies in programming and deploying algorithms via quantum circuit abstraction impede wide-range domain applications of quantum simulation. Developing better abstractions for quantum Hamiltonian simulation is highly desirable for productivity.

\vspace{3mm}
Motivated by the experimental success of simulation by designing and building specific precisely controlled quantum systems mimicking the Hamiltonian of target quantum systems \cite{zohar2015quantum, gorshkov2010two, ebadi2021quantum, yang2020observation}, programming analog quantum simulators in a Hamiltonian-oriented scheme is a promising approach to quantum applications before fault-tolerant digital quantum computers are manufactured. 
Instead of programming quantum circuits implementing quantum simulation algorithms, Hamiltonian-oriented schemes directly program Hamiltonians of analog quantum simulators to synthesize an evolution equivalent to the desired quantum system evolution.
Analog quantum simulators have native support for generating Hamiltonians, resulting in a succinct translation process to construct pulse schedules. 
Via Hamiltonian programming, complicated interactions that demand sophisticated quantum algorithms and large quantum circuits to simulate can be natively constructed and simulated on analog quantum simulators. 
We compare both schemes for quantum simulation in \fig{cH-comp} with further details. 

By breaking the quantum circuit abstraction and exposing the Hamiltonian-level programmability of modern quantum devices, resource-efficient protocols can deliver reliable solutions to quantum applications~\cite{break-abstraction} on NISQ devices, including various devices that do not support universal quantum gates, like QuEra's neutral atom devices. 

For example, using the Hamiltonian-level programmability of IBM devices, an evolution governed by $H(t)=Z_1X_2+X_2Z_3$ for time $T=1$ (formal definitions in \sec{prelim}) can be simulated by a pulse schedule with two cross-resonance pulses \cite{malekakhlagh2020first}, as illustrated in \fig{cH-comp}. Both are $280$ nanoseconds long and approximately generate Hamiltonians $Z_1X_2$ and $X_2Z_3$, respectively. Simultaneous execution of these pulses builds $H(t)$ on the IBM device, and the eventual pulse schedule is $280$ nanosecond long. As a comparison, the circuit-oriented scheme uses a circuit sequentially applying 4 CNOT gates (each requiring $264$ nanoseconds to implement) with several single qubit gates to simulate $H(t)$. It generates a pulse schedule of length $1660$ nanoseconds, around $6$ times longer. More details of this example are in \sec{cs-native}. Shortening pulse schedule duration is especially desirable because of IBM devices' short coherence time (around $100$ microseconds).

\vspace{3mm}
Although Hamiltonian-oriented approaches for quantum simulation are beneficial, there is a lack of formal abstractions and supplementary software stacks. Prior works of analog quantum simulation following Hamiltonian-oriented schemes \cite{ebadi2021quantum, yang2020observation} manually construct device-specific configurations, which are tedious, error-prone, and demanding for hardware knowledge, hence not suitable for large-scale experiments.

\begin{figure}
    \centering    
    \includegraphics[width=0.9\linewidth, trim=3cm 7cm 2.6cm 6.7cm, clip]{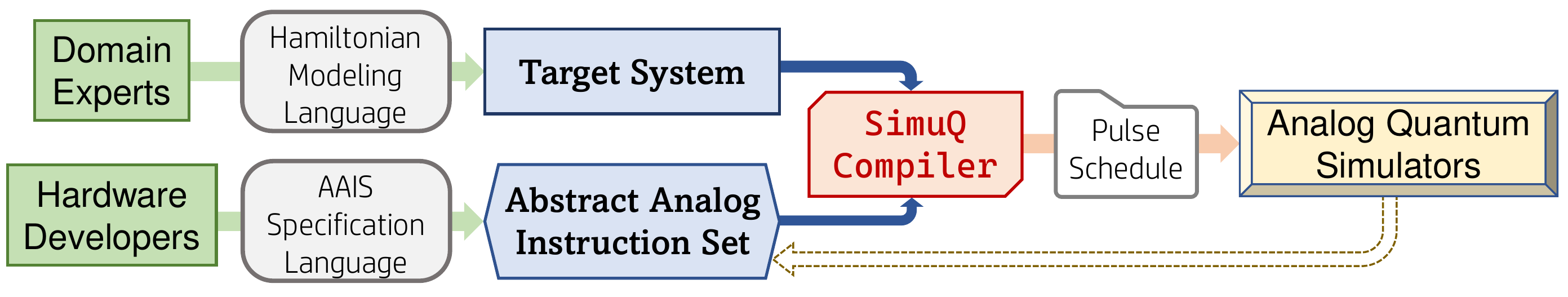}
    \vspace{-1.5em}
    \caption{The framework of SimuQ. Here abstract analog instruction sets are designed and programmed by hardware developers based on the capability of their analog quantum simulators. }
    \vspace{-2em}
    \label{fig:simuq-framework}
\end{figure}

We propose SimuQ with the first end-to-end automatic framework for quantum simulation on general analog quantum simulators, illustrated in \fig{simuq-framework}. As a result, domain experts can focus on describing the desired quantum simulation problems and leave their implementation and deployment to the automation of SimuQ. Our framework lays the foundation for large-scale applications of analog quantum simulators, paving the path for a wide range of novel and practical solutions to domain problems via quantum Hamiltonian simulation for common users.

\subsection{Challenges}

We identify three main technical challenges in building a framework to compile quantum simulation problems on analog quantum simulators: modeling the target quantum system, characterizing analog quantum simulators, and automatic compilation.

\vspace{1mm} \noindent \textbf{Modeling of quantum Hamiltonian simulation.} 
The first challenge is the lack of a scalable and user-friendly modeling language for quantum simulation.
Prior programming languages to model Hamiltonian systems are designed specifically for numerical classical simulations. They treat the sites in the quantum system as a 1-dimensional array for the convenience of constructing matrix-based mathematical objects.
One of the most popular languages, QuTiP \cite{JOHANSSON20121760-Qutip}, employs matrices of exponential sizes to represent the quantum system, resulting in poor scalability.
Another inconvenience is caused by the mandatory 1-D array labeling of the sites, like in OpenFermion \cite{mcclean2020openfermion}, Pauli IR \cite{li2022paulihedral}, and Qiskit Operator Flow \cite{aleksandrowicz2019qiskit}. Many quantum systems of interest have complicated site arrangement structures, for example, a 3-dimensional lattice, forcing users to construct the encoding of sites in their system manually.

\vspace{1mm} \noindent \textbf{Abstraction and programming of analog quantum simulators.}
The modeling of analog quantum simulators is much more challenging. 
Unlike the circuit model where the fundamental primitives are a finite number of one or two-qubit quantum gates, analog quantum simulators are usually described by continuous-time Hamiltonians on the devices with almost infinite degrees of freedom, which differ significantly among platforms~\cite{Silverio2022pulseropensource,Bloqade,Lukin-Science}. 
Moreover, complicated pulse engineering using different technologies generates various Hamiltonians with specific hardware restrictions even for one device. 
Hence a unifying and portable abstraction is in urgent demand to capture the programmability of analog quantum simulators.

\vspace{1mm} \noindent \textbf{Compilation of quantum simulations on analog quantum simulators.}
The third challenge is the lack of an automatic compilation procedure.
In the circuit-oriented scheme, the primitive gates are small-dimensional matrices. Large quantum evolution could be compiled into these gates with analytical formula (e.g., the Solovay-Kitaev theorem~\cite{kitaev1997quantum}). 
In the Hamiltonian-oriented scheme, the goal of compilation is to synthesize pulse schedules for analog quantum simulators where the Hamiltonian governing the device evolution approximately composes the target Hamiltonian $H(t)$.
This compilation process needs efficient streamlines for general analog quantum simulators of medium sizes (around 100 sites) and considers realistic hardware constraints.

\subsection{Contributions}
To the best of our knowledge, SimuQ is the first framework for programming and compiling quantum Hamiltonian simulations on heterogeneous analog quantum simulators. 
The framework tackles the above three challenges with three major components correspondingly: a new programming language for descriptions of quantum systems, a new abstraction and a corresponding programming language for characterizing the programmability of analog quantum simulators, and a compiler with several novel intermediate representations and compiler passes to deploy and execute solutions to the simulation problems on analog quantum simulators. 

\vspace{1mm} \noindent \textbf{Hamiltonian Modeling Language.} 
Without a strong design need for numerical calculations, we propose Hamiltonian Modeling Language (HML), which employs a symbolic representation treating sites as first-class objects and depicts Hamiltonians as algebraic expressions constructed via operators on the sites. 
This leads to a succinct description that remains rich enough to express many interesting quantum many-body systems. 
Users can focus on describing complicated quantum systems without tediously handcrafting encoding, reducing the cost of experimenting with new algorithm design ideas. 
Many quantum systems are programmed in HML with a few lines of code, as illustrated in \sec{cs-benchmark}. 
Beyond this, developing novel Hamiltonian-oriented quantum algorithms \cite{leng2023quantum} can benefit from HML because of the user-friendly description of the algorithms.

\vspace{1mm} \noindent \textbf{Abstract analog instruction sets and AAIS Specification Language.}
Inspired by the underlying control of these Hamiltonians, we propose a new abstraction called \emph{Abstract analog instruction set} (AAIS) to describe the functionality of heterogeneous analog devices.
Precisely, we abstract different patterns of engineered pulses as parameterized \emph{analog instructions}. 
We expose pieces of Hamiltonian in the AAIS, which are generated by analog pulses on fractions of the system and abstracted as \emph{instruction Hamiltonians} induced by instruction executions.
The Hamiltonian governing the evolution of the device at time $t$ is then the summation of instruction Hamiltonians of the instruction executions covering time $t$.

AAIS exposes the Hamiltonian-level programmability of analog quantum simulators that lies beyond circuit-level abstractions. This feature enables the programming of non-circuit-based controllable quantum devices and further exploits the capability of devices supporting quantum gates within the current hardware limits. Via Hamiltonian-level control, evolution can be simulated by a much shorter pulse duration and become more robust against device noises. 

AAIS provides a new formal computational model of quantum devices and unifies the functionality descriptions for different devices with different technologies, simplifying the transfer of quantum simulation solutions to quantum devices of multiple platforms.
These descriptions also inform theorists on what Hamiltonian-oriented quantum algorithms are realizable on near-term devices.

We propose several AAISs for QuEra, IonQ, and IBM devices. In general, AAISs should be designed by the hardware providers to expose the Hamiltonian-level control of their devices. We propose and implement AAIS Specification Language (AAIS-SL), a domain-specific language for hardware providers to depict the device programmability. We showcase how to design and program AAISs in AAIS-SL for the mentioned devices in \sec{exp-aais}.

\vspace{1mm} \noindent \textbf{SimuQ compiler.}
We propose the first compilation scheme for quantum simulation on analog quantum simulators with several new intermediate representations. We handle the synthesis of instruction executions as symbolic pattern matching inspired by the seminal work in classical analog compilation~\cite{Arco-PLDI16, 10.1145/3373376.3378449}. The instruction executions are then translated to executable pulses by resolving conflicts and reconstructing pulses using device-dependent programming languages and pulse engineering. 

To the best of our knowledge, there is no existing compilation framework for heterogeneous analog quantum simulators. Our compiler provides a feasibility demonstration of automatically compiling quantum Hamiltonian simulation on general analog quantum simulators. Although it might not be the ultimate solution, we believe our framework provides a natural and intuitive approach to the modeling and processing of necessary information in constructing executable pulses from simulation problems. The competence of our compiler is demonstrated in \sec{cs-benchmark} by showing that it can efficiently and reliably generate executable pulses for various domain applications. Pulses generated by SimuQ are executed on real devices and produce reasonable results, which has rarely been demonstrated in previous compiler works for quantum computing due to the abundance of circuit-oriented descriptions. Users can easily transport their quantum simulation experiments among different platforms and devices with our portable design of the compilation framework. It also enables the possibility of benchmarking various quantum devices on significant domain problems solvable via quantum simulation.

\vspace{2mm}

In summary, our contributions include:
\vspace{-0.2em}
\begin{itemize}[leftmargin=5.5mm]
    \item We design and implement Hamiltonian Modeling Language in \sec{hml}, a succinct DSL for describing quantum Hamiltonian simulation.
    \begin{itemize}
        \item Programs in HML are short for many important quantum systems, as shown in \sec{cs-benchmark}.
    \end{itemize}
    \item We design Abstract Analog Instruction Set as a novel abstraction of Hamiltonian-level programmability of analog quantum simulators, as illustrated in \sec{aais}. We also implement AAIS Specification Language for hardware providers to design and program AAISs.
    \begin{itemize}
        \item AAISs enable the programming of non-circuit-based quantum devices.
        \item Hamiltonian-level programming shortens pulse schedule duration and thus is more robust to device decoherence errors, with case studies detailed in \sec{cs-native} and \sec{cs-qaoa}.
    \end{itemize}
    \item We propose and implement a compiler for quantum simulation on analog quantum simulators in \sec{compiler} with new intermediate representations and compilation passes.
    \begin{itemize}
        \item SimuQ compiler enables portability among different platforms of analog quantum simulators, and generated pulses are executed on real devices, as demonstrated in \sec{cs-ising}.
        \item It efficiently compiles many significant quantum systems, as shown in \sec{cs-benchmark}.
    \end{itemize}
\end{itemize}
\vspace{-0.1em}

\vspace{1mm} \noindent \textbf{Related Works.}
There are a few Hamiltonian-level programming interfaces for analog quantum simulators, such as IBM Qiskit Pulse~\cite{openqasm3}, QuEra Bloqade~\cite{Bloqade}, and Pasqal Pulser~\cite{Silverio2022pulseropensource} developed by hardware service providers.
These interfaces are designed to represent the specific underlying quantum hardware rather than to provide a unified interface for all analog quantum simulators like AAIS. 
Computational quantum physics packages like QuTiP~\cite{JOHANSSON20121760-Qutip} support modeling and numerical calculation of quantum simulation without any compilation to quantum devices. 
Software tools for quantum Hamiltonian simulation are discussed extensively for circuit models~\cite{li2022paulihedral, schmitz2021graph, van2020circuit, POWERS2021100696, bassman2022arqtic}, while the expressiveness of the circuit abstraction limits their exploitation of analog quantum simulators. 
SimuQ's solver-based compilation is inspired by the seminal work in classical analog compilation~\cite{Arco-PLDI16,10.1145/3373376.3378449}. However, the specific abstraction and compilation technique therein is less relevant as the nature of analog quantum devices is very different from classical ones. 

%% file: 2-Example.tex
\section{Running Example} \label{sec:running}

We present a realistic but simple example to motivate our framework and showcase the methodology of our approach. Many experiments simulating the Ising model on Rydberg atom arrays are conducted in the literature to probe quantum phenomena barely tractable numerically \cite{schauss2018quantum, labuhn2016tunable}. We will introduce the mathematical description of these experiments and demonstrate how to automate the process with SimuQ's DSLs, new abstractions, and compiler.

\subsection{Quantum Preliminaries}
\label{sec:prelim}

Quantum systems consist of \emph{sites} representing physics objects like atoms, mathematically described by qubits. A \textit{qubit} (or \textit{quantum bit}) is the analogue of a classical bit in quantum computation. It is a two-level quantum-mechanical system described by the Hilbert space $\mathbb{C}^2$. 
The classical bits ``0'' and ``1'' are represented by the qubit states $\ket{0}=\left[\begin{matrix} 1 \\ 0 \end{matrix}\right]$ and $\ket{1}=\left[\begin{matrix} 0 \\ 1 \end{matrix}\right]$, and linear combinations of $\ket{0}$ and $\ket{1}$ are also valid states, forming a \emph{superpostition} of quantum states.
An $n$-qubit state is a unit vector in the Kronecker tensor product $\otimes$ of $n$ single-qubit Hilbert spaces, i.e., $\mathcal{H} = \otimes_{i=1}^n \mathbb{C}^2 \cong \mathbb{C}^{2^n}$, whose dimension is exponential in $n$. 
For an $n$ by $m$ matrix $A$ and a $p$ by $q$ matrix $B$, their Kronecker product is an $np$ by $mq$ matrix where $(A\otimes B)_{pr+u, qs+v}=A_{r, s}B_{u, v}.$ 
The \emph{complex conjugate transpose} of $\ket{\psi}$ is denoted as $\bra{\psi}=\ket{\psi}^{\dagger}$ ($\dagger$ is the Hermitian conjugate). Therefore, the \emph{inner product} of $\phi$ and $\psi$ could be written as $\braket{\phi}{\psi}$. We let $\Tr{M}$ denote the matrix trace of $M$.

The time evolution of quantum states is specified by a Hermitian matrix function $H(t)$ over the corresponding Hilbert space, known as the \textit{time-dependent Hamiltonian} of the quantum system. Typical single-site Hamiltonians include the famous \textit{Pauli matrices}:
\begingroup
\setlength\arraycolsep{3pt}
\begin{align}
\label{eqn:pauli}
    I=\left[\begin{matrix} 1 & 0 \\ 0 & 1\end{matrix}\right], ~~X=\left[\begin{matrix} 0 & 1 \\ 1 & 0\end{matrix}\right], ~~Y=\left[\begin{matrix} 0 & -i \\ i & 0\end{matrix}\right], ~~Z=\left[\begin{matrix} 1 & 0 \\ 0 & -1\end{matrix}\right].
\end{align}
\endgroup
By convention, we write $X_j$ for a multi-site Hamiltonian to indicate $I\otimes \cdots\otimes I\otimes X \otimes I\otimes\cdots\otimes I$, where the $j$-th operand is $X$. Similarly, we write $Y_j$ and $Z_j$. These notations represent operations on the $j$-th subsystem.
A product Hamiltonian $P$ is a tensor product of Pauli matrices, for example, $X\otimes I\otimes Y$, also written as $X_1Y_3$.
A multi-site Hamiltonian can be written as a linear combination of product Hamiltonians, e.g., $H=X_1X_2+2Z_2Z_3$. 
When the product Hamiltonians' coefficients are time functions, they are called time-dependent Hamiltonians, e.g., $H(t)=\cos(t)X$.
The product Hamiltonians form a complete basis of $n$-site Hamiltonians by formula
\begin{align}
    H(t)=\sum\nolimits_{P\in\{I, X, Y, Z\}^{\otimes n}}\frac{\Tr{H(t)P}}{2^n}P.
\end{align}

The time evolution of a quantum system under a time-dependent Hamiltonian $H(t)$ obeys the \emph{Schr\"odinger equation} \eq{sch}. Its solution is effectively a unitary matrix function $U(t)$ satisfying
\begin{align}
    \frac{\d}{\d t}U(t)=-iH(t)U(t).
\end{align}
If the system evolves from time $0$ with initial state $\ket{\psi(0)},$ the state at time $t$ is $\ket{\psi(t)}=U(t)\ket{\psi(0)}.$

We provide basic physics intuitions of Hamiltonian operations. Hermitian operators correspond to physics effects like the influences of magnetic fields. Scalar multiplication (e.g., $2\cdot X$) changes the effect strength. 
Additions of operators (e.g., $X_1+X_2$) represent simultaneous physics effects, e.g., the superposition of forces. Multiplications of operators (e.g., $X_1X_2$) represent the interactions across different sites, e.g., the hopping of atoms between different sites. 

A \textit{quantum measurement} extracts classical information from quantum systems. When measuring state $\ket{\phi}$, with probability $|\braket{s}{\phi}|^2,$ we obtain a classical bit-string $s$ and the quantum state $\ket{\phi}$ collapses to a classical state $\ket{s}=\ket{s_1}\otimes...\otimes\ket{s_n}$.

\subsection{Quantum Simulation of Ising Model}
\label{sec:exp-ising}

\begin{figure}
    \centering
    \begin{subfigure}[h]{0.45\linewidth}
    \centering
        \includegraphics[width=0.8\linewidth, trim=13cm 8.5cm 12.5cm 7.8cm, clip]{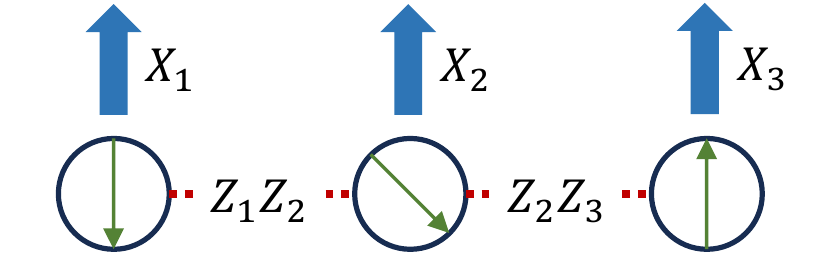}
    \vspace{-0.5em}
    \caption{An illustration of the Ising model $H_{\mathrm{Ising}}$. Here circles represent qubits, blue arrows, and red dots represent components of $H_{\mathrm{Ising}}$.}
    \label{fig:ising}
    \end{subfigure}
    \quad
    \begin{subfigure}[h]{0.5\linewidth}
        \includegraphics[width=\linewidth, trim=12.2cm 8.5cm 11cm 7cm, clip]{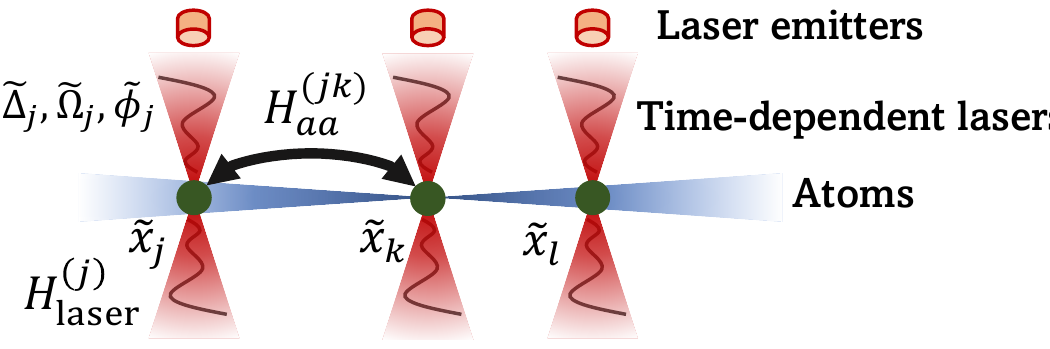}
    \vspace{-2em}
    \caption{An illustration of an ideal Rydberg device. }
    \label{fig:rydberg}
    \end{subfigure}
    \vspace{-1em}
    \caption{Illustrations of the target quantum system and the analog quantum simulator in our running example.}
\end{figure}

To understand the dynamics and properties of quantum systems, physicists have endless needs to simulate quantum systems. For example, an Ising model is mathematically expressed as
\begin{align}
    H=\sum\nolimits_{1\leq j<k\leq n}J_{jk}Z_jZ_k+\sum\nolimits_{j=1}^{n} h_jX_j
\end{align} where $J_{jk}, h_j\in\mathbb{R}$. This is a significant statistical mechanical model in the study of phase transitions of magnetic systems \cite{chakrabarti2008quantum}, with a simple example in \fig{ising}. In physics, a qubit of an Ising model represents the magnetic dipole moment of an atomic spin. $Z_j Z_k$ represents the interaction between spins $j$ and $k$, and $J_{jk}$ represents the tendency of align direction agreement between them. $X_j$ represents the effect of an external magnetic field interacting with the spins, and $h_j$ represents its strength. The evolution of a quantum system under Ising models with different parameter regimes of $J_{jk}$ and $h_j$ may characterize the magnetism of materials.
However, its simulation generally requires exponential computations for classical computers because of the exponential dimension of the Hilbert space. Instead, we consider its simulation with analog quantum simulators. Nowadays, many controllable quantum systems may be utilized for quantum simulation, and one of the most promising platforms is Rydberg atom arrays \cite{saffman2016quantum}, where neutral atoms are cooled and precisely controlled by laser beams. 

In this section, we focus on the Ising model simulation using Rydberg atom devices, whose large-scale experimental demonstrations are repeated in many laboratories \cite{labuhn2016tunable, schauss2018quantum, bernien2017probing, ebadi2021quantum}. In these demonstrations, experimentalists configure their quantum systems in a task-specific manner. The following illustration showcases these procedures, which are mostly done by manual parameter tuning. This procedure is analogous to the early-day development of classical computers before automated compilers appeared.

We consider a $3$-qubit system of the Ising model evolving for time $T$ under
\begin{align}
\label{eq:ham-ising}
    H_{\text{Ising}}=Z_1Z_2+Z_2Z_3+X_1+X_2+X_3,
\end{align}
illustrated in \fig{ising}.
We want to reproduce the evolution under $H_{\text{Ising}}$ on an \emph{ideal Rydberg device}, a simplified Rydberg atom array, illustrated in \fig{rydberg}. Mathematically, the device evolution is governed by $H_{\mathrm{Rydberg}}(\tilde{x}, \tilde{\Delta}, \tilde{\Omega}, \tilde{\phi}, t)$ where $\tilde{x}, \tilde{\Delta}, \tilde{\Omega}, \tilde{\phi}$ are configurable parameters whose details are introduced later, and $t$ is the time variable whose unit is microseconds. The goal of quantum simulation is to reproduce the evolution of $H_{\mathrm{Ising}}$ on the device by finding a configuration for $\tilde{x}, \tilde{\Delta}, \tilde{\Omega}, \tilde{\phi}$ satisfying $H_{\text{Rydberg}}(\tilde{x}, \tilde{\Delta}, \tilde{\Omega}, \tilde{\phi}, t)=H_{\text{Ising}}$, assuming the device evolution time is $T$ms.

An ideal Rydberg device contains $m$ atoms (viewed as qubits) and $m$ laser beams addressing each atom. The positions of the atoms can be configured arbitrarily on a 1-D line. We denote their coordinates as vector $\tilde{x}$ (unit: $\mu m$) and assume they will not move. A Van der Waals force acts between each pair of atoms, whose effect is described by a time-independent Hamiltonian 
\begin{align}
    H_{aa}^{(jk)}(\tilde{x}, t)=\frac{C_6}{|\tilde{x}_j-\tilde{x}_k|^6}\hat{n}_j\hat{n}_k.
\end{align}
Here $C_6\approx5.42\times 10^6 \mathrm{MHz}\cdot\mathrm{\mu m^6}$ is a real physics constant and $\hat{n}_j=(I-Z_j)/2$ is the number operator of qubit $j$. For each atom, there is a local laser beam addressing it. It contains three configurable real-function parameters $\tilde{\Delta}_j(t), \tilde{\Omega}_j(t),$ and $\tilde{\phi}_j(t)$ (unit: MHz) representing the detuning, amplitude, and phase of laser, which can be configured freely over time. It generates an effect described by
\begin{align}
    H_{\text{laser}}^{(j)}(\tilde{\Delta}_j, \tilde{\Omega}_j, \tilde{\phi}_j, t)=-\tilde{\Delta}_j(t)\hat{n}_j+\frac{\tilde{\Omega}_j(t)}{2}\left(\cos(\tilde{\phi}_j(t))X_j-\sin(\tilde{\phi}_j(t))Y_j\right).
\end{align}
Then the collective Hamiltonian governing the evolution is the summation of effects of Van der Waals forces $H_{aa}^{(jk)}$ and lasers $H_{\mathrm{laser}}^{(j)}$,
\begin{align}
    H_{\text{Rydberg}}(\tilde{x}, \tilde{\Delta}, \tilde{\Omega}, \tilde{\phi}, t)=\sum\nolimits_{1\leq j<k\leq m}H_{aa}^{(jk)}(\tilde{x}, t)+\sum\nolimits_{j=1}^m H_{\text{laser}}^{(j)}(\tilde{\Delta}_j, \tilde{\Omega}_j, \tilde{\phi}_j, t).
\end{align}

A manual way to find a device configuration is to match the coefficients of product Hamiltonians in $H_{\mathrm{Ising}}$ by configuring the parameters. Note that $Z_jZ_{j+1}$ of $H_{\text{Ising}}$ is a 2-qubit interaction which only comes from $H_{aa}^{(jk)}$. We configure $\tilde{x}_j$ accordingly by setting $\tilde{x}_j=(j-1)\times 10.52$ so that $H_{aa}^{(j(j+1))}(\tilde{x}, t)=Z_jZ_{j+1}-Z_j-Z_{j+1}+I.$ Note by setting $\tilde{x}_j$, the system has unwanted $H_{aa}^{(13)}(\tilde{x}, t)=0.016\cdot (Z_1Z_3-Z_1-Z_3+I).$
We then configure the local laser beams to create the $X_j$ terms in $H_{\text{Ising}}$ and compensate the unwanted $Z_j$ terms in $H_{aa}^{(jk)}$ by setting $\tilde{\Delta}_1(t)\equiv\tilde{\Delta}_3(t)\equiv 2.032, \tilde{\Delta}_2(t)\equiv 4, \tilde{\Omega}_j(t)\equiv 2$ and $\tilde{\phi}_j\equiv 0$. 
We can confirm our synthesis by checking $H_{\text{Ising}}-H_{\text{Rydberg}}(\tilde{x}, \tilde{\Delta}, \tilde{\Omega}, \tilde{\phi}, t)=-0.016Z_1Z_3+2.016I$. Since $2.016I$ has no measurable effects on the evolved state by quantum information analysis, The error term is $-0.016Z_1Z_3$, which is small compared to $H_{\text{Ising}}.$

\subsection{Automated Compilation by SimuQ}
\label{sec:example-ising}

SimuQ provides automation to the above procedure for analog quantum simulators by establishing a workflow via new abstractions, intermediate representations, and compilation passes designed explicitly for analog compilation of quantum simulation. We illustrate how to program and compile $H_{\text{Ising}}$ on the ideal Rydberg device in SimuQ, with a glimpse of our new abstractions. 

\begin{figure*}[b]
    \begin{adjustbox}{minipage=0.34\linewidth, scale=0.85}
\hspace{1em}
    \begin{subfigure}[t]{\linewidth}
    \begin{pythonnum}
Ising = QSystem()
q = [Qubit(Ising) 
        for i in range(N)]
h = 0
for i in range(N) :
    h += q[i].X
for i in range(N - 1) :
    h += q[i].Z * q[i+1].Z
Ising.add_evolution(h, T)
    \end{pythonnum}
    \vspace{-0.5em}
    \caption{An evolution governed by $H_{\mathrm{Ising}}$ \eq{ham-ising} programmed in HML. Here $N=3$ is the number of sites. $T=1$ is the evolution time.}
    \label{fig:ising-prog}
    \end{subfigure}
    \end{adjustbox}
    \qquad \qquad
    \begin{adjustbox}{minipage=0.65\linewidth, scale=0.75}
    \begin{subfigure}[t]{\linewidth}
    \begin{pythonnum}
Rydberg = QMachine()
q = [qubit(Rydberg) for i in range(N)]
n = [(q[i].I - q[i].Z) / 2 for i in range(N)]
for i in range(N) :
    $\eta$ =  Rydberg.add_instruction()
    add_lVar = $\eta$.add_local_variable
    $\Delta$, $\Omega$, $\phi$ = $~$add_lVar(), add_lVar(), add_lVar()
    X_Y = cos($\phi$) *  q[i].X - sin($\phi$) *  q[i].Y
    $\eta$.set_ham(-$\Delta$ *  n[i] + $\Omega$ /  2 * X_Y)
x = [Rydberg.add_global_variable() for i in range(N)]
h = 0
for i in range(N) :
    for j in range(i) :
        h += (C / (x[i] - x[j])**6) * n[i] * n[j]
Rydberg.set_sys_ham(h)
    \end{pythonnum}
    \vspace{-0.5em}
    \caption{The Rydberg AAIS programmed in AAIS-SL. $N=3$ is the number of sites. $C$ is the Rydberg interaction constant.}
    \label{fig:aais-rydberg}
    \end{subfigure}
    \end{adjustbox}
    \vspace{-1em}
    \caption{Examples of HML and AAIS specification language implemented in Python. }
    \label{fig:python-code}
\end{figure*}

\paragraph{Programming an Ising evolution}

Firstly, we program $H_{\text{Ising}}$ in HML with an implementation in Python as in \fig{ising-prog}.
The first step is to declare a quantum system \pyth{Ising} (Line 1) and three sites (qubit) belonging to it (Line 2-3). By storing the sites in a list, we refer to the $j$-th site of the system by \pyth{q[j]}. Then we construct $H_{\text{Ising}}$'s terms one by one and store them in \pyth{h} (Line 4-8). Here we program the terms as an expression containing operators on the sites, e.g., $X_j$ as \pyth{q[j].X} and $Z_jZ_{j+1}$ as \pyth{q[j].Z*q[j+1].Z}. We then let the system evolve under \pyth{h} for time \pyth{T} (Line 9). 

\paragraph{Characterizing ideal Rydberg devices}

We propose a Rydberg AAIS to characterize the programmability of ideal Rydberg devices. 
An implementation is in \fig{aais-rydberg}.

The program starts with declaring the quantum device (Line 1) and its sites (Line 2). We can construct the number operators $\hat{n}_j$ and store them in \pyth{n} (Line 3).

We propose \emph{analog instructions} to characterize the effects produced and configured on the device over time. An instruction execution generates an instruction Hamiltonian that adds up to the total Hamiltonian governing the system. 
In the Rydberg AAIS, we design instructions $\eta_j$ to model the effects of the laser beam pulse signals (Line 5). Executing $\eta_j$ generates an instruction Hamiltonian $\ssem{\eta_j}(\Delta_j, \Omega_j, \phi_j)=\Delta_j\hat{n}_j+\Omega_j/2(\cos(\phi_j)X_j-\sin(\phi_j)Y_j)$, where $\Delta_j, \Omega_j,$ and $\phi_j$ are \emph{local variables} belonging to $\eta_j$ (Line 6-9). When executing $\eta_j$, one may specify a valuation $\vec{b}$ and execution starting and ending time $\tau_s, \tau_e$ to induce a Hamiltonian $\ssem{\eta_j}(\vec{b})$ on the device in time interval $[\tau_s, \tau_e)$. 

Instructions can be executed simultaneously to create an evolution under their collective effects, mathematically expressed as a summation of instruction Hamiltonians. For example, we can simultaneously switch on the laser beams addressing atoms 1 and 2 with configuration $\vec{b}_1$ and $\vec{b}_2$ and switch off others, generating a Hamiltonian $\ssem{\eta_1}(\vec{b}_1)+\ssem{\eta_2}(\vec{b}_2).$ 

Besides the instructions executed over time, analog quantum simulators may also have inherent effects, like the atom-atom interactions in the Rydberg atom devices. We declare a \emph{system Hamiltonian} with \emph{global variables} for them. Let $x_j$ be the global variables representing the position of the atoms (Line 10). Then collective Van der Waals force $\sum_{1\leq j<k\leq m}H_{aa}^{(jk)}(x, t)=\sum_{1\leq j<k\leq m}C_6/|x_j-x_k|^6\hat{n}_j\hat{n}_k$ is characterized as the system Hamiltonian $H_{\mathrm{sys}}(\vec{x})$ (Line 11-15).

Overall, the Hamiltonian governing an ideal Rydberg device at time $t$ is
\begin{align}
    H_{IRD}(t)=H_{\text{sys}}(\vec{x})+\sum\nolimits_{(\eta_j, \vec{b}_j)\in C_t} \ssem{\eta_j}(\vec{b}_j),
\end{align}
where $C_t$ contains the active instruction executions at time $t$ and their variable valuations.

\paragraph{Synethsizing $H_{\text{Ising}}$ on ideal Rydberg devices}
The SimuQ compiler automatically synthesizes a target Hamiltonian with an AAIS and generates executable pulses for devices. We go through the compilation steps on a 3-atom ideal Rydberg device, creating a configuration satisfying $H_{IRD}(t)=H_{\mathrm{Ising}}.$

The first step is to find a site layout between the Hilbert space of $H_{\text{Ising}}$ and the Hilbert space of ideal Rydberg devices. A trivial layout that maps the $j$-th site of $H_{\text{Ising}}$ to the $j$-th atom of the ideal Rydberg device suffices since the atoms are homogeneous.

\begin{figure}
    \centering
    \includegraphics[width=0.9\linewidth, trim = 2.5cm 6cm 4cm 4.5cm, clip]{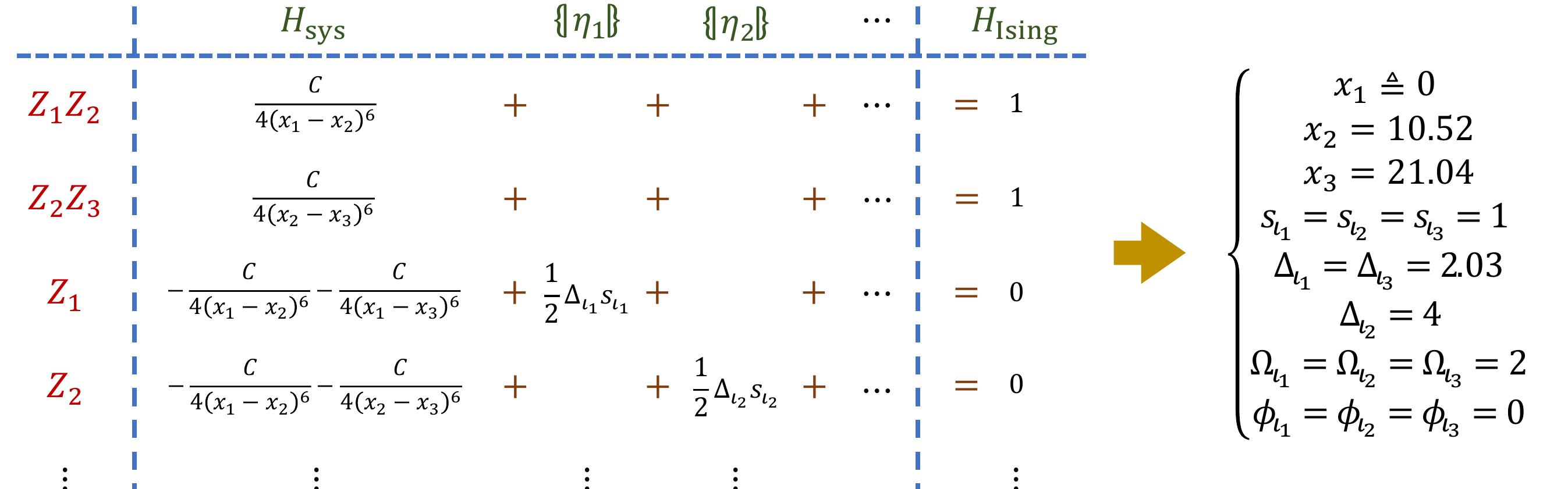}
    \caption{On the left is the equation system to synthesize $H_{\mathrm{Ising}}$ using the AAIS for the ideal Rydberg device. On the right is an approximate solution to it. It can be further interpreted as a pulse schedule in Bloqade. }
    \label{fig:coef-eq}
\end{figure}

For simplicity, here we assume the on-device evolution time is the target evolution time $T$. We then synthesize $H_{\mathrm{Ising}}$ by matching the coefficients of its product Hamiltonians. 
For a product Hamiltonian $P$, let $H[P]$ be the coefficient of $P$ in Hamiltonian $H$ and $\ssem{\eta_j}[P]$ be the coefficient function of $P$ in $\ssem{\eta_j}$.
We take product Hamiltonian $Z_1$ as an example, whose coefficient is $H_{\text{Ising}}[Z_1]=0$. $Z_1$ has non-zero coefficient functions in $\ssem{\eta_1}$ and the system Hamiltonian $H_{\mathrm{sys}}$
\begin{align}
    \ssem{\eta_1}[Z_1](\Delta_1, \Omega_1, \phi_1)=\frac{\Delta_1}{2}, \qquad H_{\mathrm{sys}}[Z_1](\vec{x})=-\frac{C_6}{4|x_1-x_2|^6}-\frac{C_6}{4|x_1-x_3|^6}.
\end{align} 
We want a set of instruction executions letting the coefficient of $Z_1$ be $0$.
Since instruction $\eta_1$ is optionally executed, an indicator variable $s_1\in\{0, 1\}$ is declared to represent whether $\eta_1$ is executed.
Then we establish an equation 
\begin{align}
    H_{\mathrm{sys}}[Z_1]+\ssem{\eta_1}[Z_1]\cdot s_1=H_{\mathrm{Ising}}[Z_1] \quad \Leftrightarrow \quad -\frac{C_6}{4|x_1-x_2|^6}-\frac{C_6}{4|x_1-x_3|^6}+\frac{\Delta_1}{2}s_1=0.
\end{align}
In general, we declare $s_j\in\{0, 1\}$ for each instruction $\eta_j$ and establish an equation system
\begin{align}
    \forall P\neq I, \quad H_{\text{sys}}[P]+\sum\nolimits_j \ssem{\eta_j}[P]\cdot s_j=H_{\text{Ising}}[P]
\end{align}
by enumerating every $P$ to match all coefficients of product Hamiltonians in $H_{\text{Ising}}$. \fig{coef-eq} shows other established equations, and the full equation system is in \app{eqset}.

We employ a numerical solver to search for a solution to the non-linear mixed binary equation system. An approximate solution to the equation system is displayed in \fig{coef-eq}. We interpret the solution as an instruction schedule: it specifies a collection of instruction executions according to the solutions to $s_j$ and local variables. The solution in \fig{coef-eq} can then be interpreted: set the positions of atoms at $x=[0, 10.52, 21.04]\mu m$, set laser beams configuration $(\Delta(t), \Omega(t), \phi(t))\equiv(2.032, 2, 0)$ for atom $1$ and $3$ and $(\Delta(t), \Omega(t), \phi(t))\equiv(4, 2, 0)$ for atom $2$, and evolve the system for $T$ms. These configurations can be translated to a Bloqade or Braket program to execute on QuEra devices.

With the above procedure, we successfully simulate the evolution under $H_{\text{Ising}}$ on the ideal Rydberg device with the help of SimuQ. In practice, hardware providers design AAIS and implement the analog instructions for their specific devices. Front-end users only need to program $H_{\text{Ising}}$ and employ SimuQ to generate executable code to send to backend devices. SimuQ breaks the knowledge barriers for frontend users to exploit analog quantum simulators easily. The following sections will explicate SimuQ components and technical details. 

%% file: 3-DSL.tex
\section{Domain-Specific Languages} \label{sec:dsl}

SimuQ is the first framework to tackle quantum simulation with Hamiltonian-level compilation to analog quantum simulators. It includes a collection of novel abstractions and domain-specific languages (DSL). We propose two DSLs in SimuQ: Hamiltonian Modeling Language (HML) for front-end users to depict their target quantum systems and AAIS Specification Language (AAIS-SL) to specify analog abstract instruction sets (AAISs) of analog quantum simulators.

\subsection{Hamiltonian Modeling Language}
\label{sec:hml}

HML is a DSL designed to describe the physical structure of many-body quantum systems that introduces many abstractions, including quantum sites and site-based representations of Hamiltonians. We implement this language in Python, with its abstract syntax and denotational semantics formally defined in \fig{HML}.

\begin{figure}[t]
\begin{minipage}{0.5\linewidth}
\scalebox{0.9}{
\begin{subfigure}[c]{\linewidth}
\begin{align*}
    & A\in \text{Site}, & r\in\mathbb{R}, \tau \in \mathbb{R}^+
\end{align*}
\begin{align*}
    R &\in \text{Operator} &::=~~& I ~|~ X ~|~ Y ~|~ Z \\
    S &\in \text{Scalar} &::=~~& S_1+S_2 ~|~ S_1\cdot S_2 ~|~ S_1 - S_2 ~|~ S_1 / S_2 \\
                          & & &  ~|~ \text{exp}(S) ~|~ \cos(S) ~|~ \sin(S) ~|~ r \\
    M &\in \text{Hermitian} &::=~~& M_1 + M_2 ~|~ M_1 \cdot M_2 ~|~ S \cdot M ~|~ A.R \\
    E &\in \text{Evolution} &::=~~& \nil ~|~ (M, \tau); E
\end{align*}
\vspace{-1.5em}
\caption{Abstract syntax of HML.}
\label{fig:syntax-HML}
\end{subfigure}
}
\end{minipage}
~
\begin{minipage}{0.45\linewidth}
\scalebox{0.9}{
\begin{subfigure}[t]{\linewidth}
\begin{align*}
    h_{A.R}&=R_{A}\\
    h_{S\cdot M}&=\mathrm{eval}(S)\cdot h_M, \\
    h_{M_1+M_2}&=h_{M_1}+h_{M_2}, \\
    h_{M_1\cdot M_2}&=h_{M_1}\cdot h_{M_2},\\
    \sem{\nil} &= I, \\
    \sem{(M, \tau ); E} &= \sem{E}\cdot e^{-i\tau h_{M}}.
\end{align*}
\vspace{-1.5em}
\caption{Semantics of HML. }
\label{fig:sem-HML}
\end{subfigure}
}
\end{minipage}
\vspace{-1em}
\caption{Syntax and denotational semantics of HML. Here Site contains system sites. $h_M$ translates to the Hermitian matrix described by $M$. $R_{A}$ is a Hermitian matrix where operator $R$ applies to site $A$ and $I$ applies to other sites. Function $\mathrm{eval}$ evaluates scalar expression $S$ to a real number.}
\label{fig:HML}
\end{figure}

\paragraph{Abstract Syntax of HML}
The first-class objects in HML are \emph{sites of quantum systems}. A site is an abstraction for any quantized 2-level physical entity, like atoms with two energy levels, whose mathematical description is a qubit.
In HML, site identifiers are collected in a set Site, each representing a site of the system. 
Four operators, $I, X, Y$, and $Z$, are defined to represent the Pauli operators, and they are \emph{site operators}. We denote the $X$ operator of qubit $q$ as $q.X$ and other operators similarly. 

A time-independent Hamiltonian is effectively a Hermitian matrix programmed by algebraic expressions. The basic elements are site operators $A.R$. Expressions for Hermitian are constructed using site operators and scalar expressions, consisting of common matrix operations and scalar operations.
An evolution $E$ in HML is a sequence of pairs $(M, \tau)$, representing a sequential evolution where each segment is governed by a time-independent Hamiltonian $h_M$ and for time $\tau$. 

\begin{remark}
    Beyond sites representing qubits, sites representing fermionic and bosonic modes can be defined together with their annihilation and creation operators. These are characterized by different types of sites in our implementation. Each type of site contains specific site operators, and the operator algebras are symbolically implemented. We omit formal discussions of them in this paper for simplicity. 
\end{remark}

\begin{remark}
    HML can generally deal with Hamiltonians with continuous-time coefficients by introducing an additional identifier $t$ in scalars. We choose sequences of time-independent evolution for numerical convenience in the compilation stage and leave this possibility for the future.
\end{remark}

\paragraph{Semantics of HML}
The denotational semantics of a program $E$ in HML is interpreted as a unitary matrix by $\sem{E}$ in \fig{sem-HML}. We let $h_M$ translate program $M$ into Hermitian matrices by evaluating the expressions. Then $\sem{E}$ is the product of unitary matrices $e^{-i \tau h_M}$, each representing the solution to the Schr\"odinger equation under $H(t)=h_M$ for time duration $\tau$. This is the solution to the Schr\"odinger equation governed by the piecewise-constant Hamiltonian programmed in $E$. 

\paragraph{Implementation of HML}

We implement HML in Python to ensure accessibility to physicists and other common users. For a quantum system, we store the sites in a list. A product Hamiltonian $P$ is then stored as a list of site operators using the same order of the site list. We also employ a Python dictionary to store a time-independent Hamiltonian $H$ where the key-value pairs are made of a product Hamiltonian $P$ and its coefficient denoted by $H[P]$. Mathematically, $H[P]=\Tr{H\cdot P}$. We only store those $P$ with non-zero $H[P]$ to compactly store Hamiltonians. For example, $H_{\text{Ising}}$ in \sec{example-ising} is represented by a dictionary $\left\{Z_1Z_2: 1, ~ Z_2Z_3: 1, ~ X_1:1, ~ X_2:1, ~X_3:1\right\}.$

To deal with the algebraic operations of Hermitian matrices, we symbolically implement an algebraic group for site operators (the Pauli group), and then Hermitian expressions are evaluated accordingly. For example, $H_1+H_2$ is effectively implemented by enumerating $P$ appearing in the keys of $H_1$'s and $H_2$'s dictionary, and construct $(H_1+H_2)[P]=H_1[P]+H_2[P].$ Another example is multiplication, where $H_1\cdot H_2$ is implemented by enumerating $P_j$ in $H_j$'s dictionary keys. Since the site operators of different sites commute and those of the same sites are in a finite group, $P_1\cdot P_2$ is a product Hamiltonian $P$ with an additional scalar multiplier $p$ (i.e., $(X_1X_2)\cdot (Y_1Y_2)=-1\cdot Z_1Z_2$). We add $p\cdot H_1[P_1]\cdot H_2[P_2]$ to the coefficient $(H_1\cdot H_2)[P]$. Then we represent the evolution $E$ as a list of tuples $(H, \tau)$ encompassing Hermitian matrix $H$ and the evolution time $\tau$ of an evolution segment.

\paragraph{Input Discretization Error}
In many-body physics systems, Hamiltonians are commonly continuous, taking form $H_{\mathrm{tar}}(t)=\sum_{k=1}^K\alpha_k(t)H_k$. In HML, these Hamiltonians are discretized into a series of piecewise time-independent Hamiltonians in the input. Let the evolution duration be $T$ and the discretization number be $D$. We discretize $H_{\mathrm{tar}}(t)$ over time steps $\{t_d\}_{d=1}^D$ where $0<t_1<...<t_D<T$ and use the left endpoint of each interval as its approximation. Formally, $H_{\mathrm{tar}}(t)$ is approximated by 
\begin{align}
    \tilde{H}(t)&=\sum\nolimits_{k=1}^K\tilde{\alpha}_k(t)H_k,\qquad \tilde{\alpha}_k(t)=\sum\nolimits_{d=1}^{D}\alpha_{k}(t_d)\mathds{1}_{[t_d, t_{d+1})}(t),
\end{align} 
where $\mathds{1}_{[a, b)}$ is the indicator function of set $[a, b)$. We assume $\norm{H_k}=1$ where $\norm{\cdot}$ is the spectral norm of matrices, $\alpha_k(t)$ are piecewise $M$-Lipschitz functions, and $\{t_d\}_{d=1}^D$ include all partitioning points of the piecewise Lipschitz coefficients $\alpha_k(t)$. Then we can derive the error bound induced by discretization by the following lemma.

\begin{lemma}[\cite{nielsen2002quantum}]
\label{lem:discretize}
The difference between the unitary $U(T)$ of evolution under $H_{\mathrm{tar}}(t)$ for duration $T$ and the unitary $\tilde{U}(T)$ of evolution under $\tilde{H}(t)$ is bounded by 
\begin{align}
    \norm{U(T)-\tilde{U}(T)}\leq C_1 D^{-1} M K T^2.
\end{align}
Here $C_1>0$ is a constant, $D$ is the discretization number, $K$ is the number of terms in $H_{\mathrm{tar}}(t)$, and $L$ is the Lipschitz constant for $\alpha_k(t)$.
\end{lemma}

This lemma shows that when we increase the discretization number $D$, the evolution error in the approximation can be arbitrarily small, justifying the discretization. The proof is routine in quantum information and hence omitted.

\subsection{Abstract Analog Instruction Set and AAIS Specification Language}
\label{sec:aais}

\begin{figure}[t]
\begin{minipage}{0.53\linewidth}
\scalebox{0.9}{
\begin{subfigure}[c]{\linewidth}
\begin{align*}
    & A\in \text{Site}, ~~ v[q]\in \text{Var}^q \text{ for } q\in\{L, G\}, ~~ r\in\mathbb{R}
\end{align*}
\begin{align*}
    R &\in \text{Operator} &::=~~& I ~|~ X ~|~ Y ~|~ Z \\
    S^q &\in \text{Para. Scalar}^q &::=~~& S^q_1+S^q_2 ~|~ S^q_1\cdot S^q_2 ~|~ S^q_1 - S^q_2 ~|~ S^q_1 / S^q_2\\
                               & & & ~|~ \text{exp}(S^q)~|~ \cos(S^q) ~|~ \sin(S^q) ~|~ r ~|~ v[q] \\
    M^q &\in \text{Para. Herm.}^q &::=~~& M^q_1 + M^q_2 ~|~ M^q_1 \cdot M^q_2 ~|~ S^q \cdot M^q ~|~ A.R \\
    D &\in \text{Device} &::=~~& M^G ~|~ M^L; D
\end{align*}
\vspace{-1.5em}
\caption{Syntax of AAIS specification language. }
\end{subfigure}
}
\end{minipage}
\hspace{2em}
\begin{minipage}{0.4\linewidth}
\scalebox{0.9}{
\begin{subfigure}[t]{\linewidth}
\begin{align*}
    h_{A.R}&=R_{A},\\
    h_{S^q\cdot M^q}&=\overline{\mathrm{eval}}(S^q)\cdot h_{M^q}, \\
    h_{M^q_1+M^q_2}&=h_{M^q_1}+h_{M^q_2}, \\
    h_{M^q_1\cdot M^q_2}&=h_{M^q_1}\cdot h_{M^q_2},\\
    \ssem{M^G}&=h_{M^G}, \\
    \ssem{M^L; D}&=h_{M^L}; \ssem{D}
\end{align*}
\vspace{-1.5em}
\caption{Semantics of AAIS programs. }
\label{fig:sem-aais}
\end{subfigure}
}
\end{minipage}
\vspace{-1em}
\caption{Abstract syntax and denotational semantics of AAIS-SL. Here Site contains the sites of the device. 
$\text{Var}^L$ and $\text{Var}^G$ contain the local and global variables correspondingly. $\overline{\mathrm{eval}}(S)$ evaluates $S$ as a real function. }
\label{fig:AAIS}
\end{figure}

\subsubsection{Abstract Analog Instruction Set}

An AAIS conveys the functionality of an analog quantum simulator in the form of instructions and system Hamiltonians, including necessary device information for synthesizing target quantum systems. 

We present the AAIS design and their physics correspondences in \tab{AAIS-comp}. An \emph{analog instruction} $\eta$ of an AAIS contains configurable parameters $\vec{v}$ and generates an \emph{instruction Hamiltonian} $H_{\eta}(\vec{v})$ on the device when executed. These parameters are \emph{local variables} of $\eta$, modeling the device parameters that can change over time. 
The instruction Hamiltonian $H_{\eta}(\vec{v})$ takes the following form
where $u_P(\vec{v})$ is a real function depending on the local variables $\vec{v}$:
\begin{align}
\label{eq:ins-ham}
    H_{\eta}(\vec{v})=\sum\nolimits_{P} u_P(\vec{v})\cdot P.
\end{align}

Additionally, a \emph{system Hamiltonian} $H_{\mathrm{sys}}(\vec{v}_{\mathrm{glob}})$ with a similar form of \eq{ins-ham} applies an always-on effect on the device. A vector $\vec{v}_{\mathrm{glob}}$ of time-independent configurable parameters, called \emph{global variables}, belongs to it. These global variables are configured before executing any instructions and stay unchanged during the execution.

\begin{table}
    \scalebox{0.85}{
    \centering
        \begin{tabular}{|c|c|c|c|c|}
        \hline
             \textbf{Physics} & Signal carriers & Pulse signals & Signal effects & Device evolution \\
        \hline
             \textbf{Rydberg devices} & Laser emitters & Time-dependent lasers & $H_{\mathrm{laser}}^{(j)}$ & Obeys $H_{\mathrm{Rydberg}}(t)$ \\
        \hline
             \textbf{AAIS} & Signal lines & Instructions & Instruction Hamiltonians & Total Hamiltonian \\
        \hline
        \end{tabular}
    }
        \caption{Comparison among physics concepts, Rydberg devices instances, and AAIS abstraction designs. }
        \label{tab:AAIS-comp}
\vspace{-2em}
\end{table}

\subsubsection{AAIS Specification Language}

To specify AAISs with programs, we propose and implement AAIS-SL and present its abstract syntax and denotational semantics in \fig{AAIS}.

\paragraph{Abstract Syntax of AAIS-SL}

To characterize the Hamiltonians of instructions, sites are declared with identifiers stored in a set Site, and site operators are defined as objects of sites by default.

Compared to HML, the major difference in the syntax is variables. Two types of variables whose identifiers are stored in $\text{Var}^G$ and $\text{Var}^L$ represent global variables and local variables, respectively. They are terms in parameterized scalars and consist of parameterized Hermitians. Then an AAIS for a device is effectively a collection of instruction Hamiltonians as parameterized Hermitian matrices, along with the system Hamiltonian. 

\paragraph{Denotational Semantics of AAIS-SL}

We interpret an AAIS $D$ characterizing a device as a list of instructions along with the system Hamiltonian. Similar to the HML semantics, we employ a translation $h$ for expressions $S$ to obtain parameterized Hermitians. Function $\overline{\mathrm{eval}}$ evaluates a parameterized scalar expression $S$ as a real function taking a valuation of variables and outputting a real number. Hence $h$ translates parameterized Hermitian expressions to Hamiltonians in the form of \eq{ins-ham}. Without ambiguity, we use $\ssem{\eta}(\vec{v})$ to represent the instruction Hamiltonian of $\eta$.

\paragraph{Implementation of AAIS-SL}

We also provide a Python implementation of AAIS-SL. We store sites and Hermitian matrices similarly to the implementation of HML. The difference is that instead of storing real numbers as coefficients, we store Python functions taking global variable and local variable valuations as inputs. We build function algebraic operations (i.e., $(f_1+f_2)(x)=f_1(x)+f_2(x)$ and $(f_1\cdot f_2)(x)=f_1(x)\cdot f_2(x)$) to deal with expressions and establish the parameterized Hermitian matrix expressions. As described in \fig{AAIS}, an AAIS is effectively represented by a system Hamiltonian and a list of instruction Hamiltonians. 

\subsubsection{Examples of AAIS}
\label{sec:exp-aais}

Through AAISs, we provide a general framework to characterize the programmability of analog quantum simulators. 
Here we show how we design AAISs for QuEra, IonQ, and IBM devices. The design of AAIS abstraction pursues a balance between expressiveness and implementation hardness on real devices: to simulate more complicated quantum systems, more complicated instructions are needed, requiring more advanced technologies in their implementation.

\paragraph{Rydberg AAIS} The Rydberg AAIS designed for ideal Rydberg atom devices is introduced in \sec{example-ising}. However, current QuEra devices do not support local laser addressing, meaning only a global laser interacts with every atom simultaneously. We use a variant for QuEra devices, called the global Rydberg AAIS, where there is only one instruction $\eta$ with the instruction Hamiltonian 
\begin{align}
    \ssem{\eta}(\Delta, \Omega, \phi)=-\Delta\sum\nolimits_{j=1}^m \hat{n}_j+\frac{\Omega}{2}\sum\nolimits_{j=1}^m (\cos(\phi)X_j-\sin(\phi)Y_j).
\end{align}

\paragraph{Heisenberg AAIS} The IonQ and IBM devices, though using different platforms, share similar capabilities for constructing interactions. For both platforms, the Heisenberg AAIS is designed and implemented, which contains 1-site instructions $\eta_{j, P}$ and 2-site instructions $\eta_{j, k, PP}$ for $j, k\in\{1, ..., n\}$ and $P\in\{X, Y, Z\}$, where $n$ is the number of sites. Each instruction possesses one local variable, and their instruction Hamiltonians are:
\begin{align}
    \ssem{\eta_{j, P}}(a)=a\cdot P_j, \qquad 
    \ssem{\eta_{j, k, PP}}(a)=a\cdot P_jP_k.
\end{align}
Here, the 1-site instructions $\eta_{j, P}$ are defined for every site in the system, and the 2-site instructions $\eta_{j, k, PP}$ are only defined when $(j, k)\in E$ for an undirected connectivity graph $E$ representing the connectivity of the detailed device. For ion trap devices, $E$ is a complete graph with an edge between each site pair. Superconducting devices typically have limited connectivity, and we let $E$ be the connectivity graph of the IBM devices. 

The Heisenberg AAIS can simulate a family of Heisenberg models \cite{auerbach1998interacting} covering the Ising models.
A variant of the Heisenberg AAIS called the 2-Pauli AAIS extends the 2-site interactions to $P_jQ_k$ interactions for $P, Q\in\{X, Y, Z\}$, is capable of simulating more quantum systems, and is realizable on IonQ and IBM devices with specific connectivity. 

\paragraph{IBM-Native AAIS} Besides the Heisenberg AAIS, for the IBM devices, we can also model their native effects in an IBM-native AAIS. Its 2-site instructions are $\eta_{j, k, CR}$ for $(j, k)\in E$ where 
\begin{align}
    \ssem{\eta_{j, k, CR}}(\Omega)=\omega_{ZX}\Omega Z_jX_k+\omega_{ZZ}Z_jZ_k+\omega_{IX}\Omega X_k+\omega_{ZI}\Omega^2 Z_j,
\end{align}
where $\omega_{ZX}, \omega_{ZZ}, \omega_{IX},$ and $\omega_{ZI}$ are device-dependent constants. Instruction $\eta_{j, k, CR}$ and $\eta_{l, k, CR}$ can be simultaneously executed on IBM devices because of platform features. However, since it contains multiple terms with limited freedom of control, only a few quantum systems can be directly simulated by the IBM-native AAIS. For the systems that can be simulated, a much shorter pulse duration can be produced. A more detailed analysis is in \sec{cs-native}.

%% file: 4-Compiler.tex
\section{Intermediate Representations and Compilation} \label{sec:compiler}

Compiling a target quantum system to an analog quantum simulator is computationally hard in most cases, especially when we aim at a general framework. In this section, we build the first compiler for quantum simulation on general analog quantum simulators and several novel intermediate representations to conquer various challenges in the overall compilation. 

\begin{figure}
    \centering
    \includegraphics[width=0.9\linewidth, trim=1.5cm 4.5cm 2cm 6cm, clip]{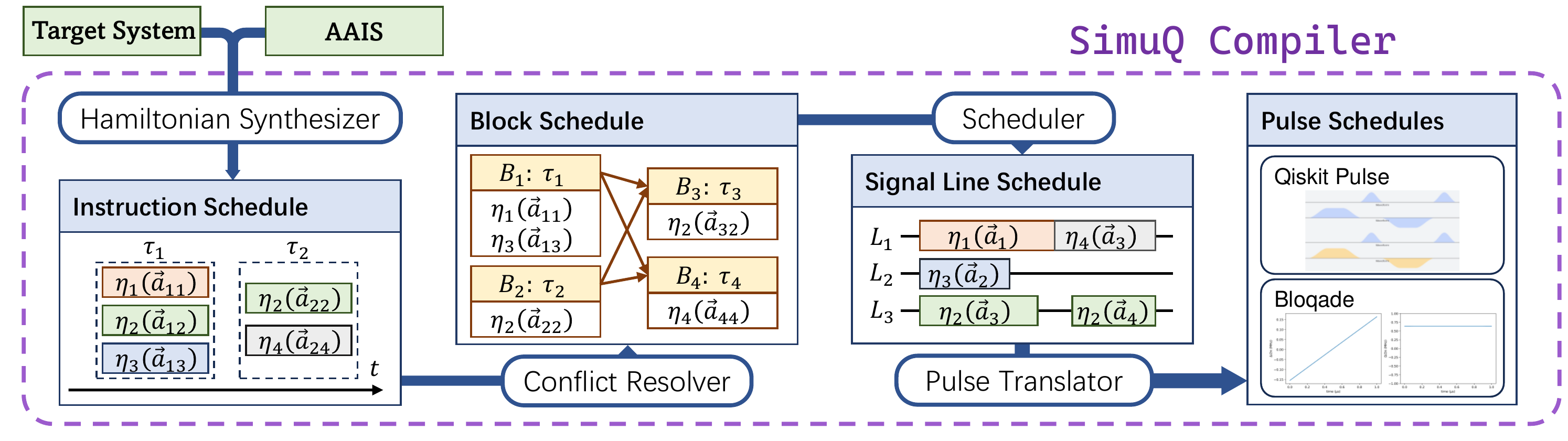}
    \vspace{-1em}
    \caption{An illustration of the SimuQ compilation process. }
    \vspace{-1em}
    \label{fig:compiler}
\end{figure}

The overall compilation workflow is presented in \fig{compiler}. Since this is the first exploration of compilation to heterogeneous analog quantum simulators, our proposal intuitively decomposes the problem into several natural sub-problems that are rarely encountered in prior works and applies straightforward solutions to each step. Much space for optimizing our workflow within the scope of our approach is left for future work, which is discussed in \sec{future-work}.

\subsection{Instruction Schedules and Hamiltonian Synthesizer}

The first intermediate representation is instruction schedules that describe the execution of instructions on the device. We will also introduce a Hamiltonian synthesizer to create an instruction schedule that simulates a target quantum system.

\subsubsection{Instruction Schedules}

An instruction execution $(\eta, \vec{a}, \tau_s, \tau_e)$ specifies an instruction $\eta$, a valuation $\vec{v}\mapsto\vec{a}$ of $\eta$'s local variables, and evolution starting time $\tau_s$ and ending time $\tau_e$. It applies a Hamiltonian $H_{\eta}(\vec{a})$ to the device during $[\tau_s, \tau_e)$. In later cases when the absolute starting time and ending time are unimportant, we also use duration $\tau_d=\tau_e-\tau_s$ in instruction executions.

An instruction schedule includes a valuation $\vec{g}$ to the global variables and a set of instruction executions $\{(\eta_j, \vec{a}_j, \tau_{s, j}, \tau_{e, j})\}$. At the time $t$, the instruction executions satisfying $\tau_{s, j}\leq t<\tau_{e, j}$ generate effects on the device. Executing the instruction schedule evolves the device, governed by:
\begin{align}
    H(t)=H_{\mathrm{sys}}(\vec{g})+\sum\nolimits_{j: \tau_{s, j}\leq t<\tau_{e, j}}H_{\eta_j}(\vec{a}_j).
\end{align}

We use a more succinct representation of the instruction schedules generated by our Hamiltonian synthesizer. We characterize the set of instruction executions as
a list $\S=[(C_j, \tau_j)]_{j=1}^{m}$ where $C_j=\{(\eta_{jk}, \vec{a}_{jk})\}_k$. $\S$ denotes a sequential evolution of simultaneous instruction executions in $C_j$ for time duration $\tau_j$. Let $T_j=\sum_{k\leq j}\tau_j$ and assume $T_0=0$. The absolute starting and ending time of instruction execution $(\eta_{jk}, \vec{a}_{jk})\in C_j$ are then $T_{j-1}$ and $T_j$. Mathematically, the Hamiltonian $H(t)$ governing the evolution of the device at time $t\in[T_{j-1}, T_{j})$ is $H(t)=H_{\text{sys}}(\vec{g})+\sum_{k}\ssem{\eta_{jk}}(\vec{a}_{jk}).$
As a solution to the Schr\"odinger equation, the execution of instruction schedule $(\S, \vec{g})$ results in an evolution of the device described by a unitary matrix 
\begin{align}
    U(T_m)=\prod\nolimits_{j=m}^1e^{-i\tau_j\left(H_{\text{sys}}(\vec{g})+\sum_k\ssem{\eta_{jk}}(\vec{a}_{jk})\right)}.
\end{align}

\subsubsection{Quantum Simulation by Executing Instruction Schedules}

We formally define the task of compiling quantum simulations to a quantum device described by an AAIS. Consider a target quantum system described by a Hamiltonian $H_{\text{tar}}(t)$ and evolution time interval $[0, T)$. Compilation of a quantum simulation asks for a site layout $L$ and an instruction schedule $(\S, \vec{g})$. A site layout $L$ is an injective mapping from each site in the target system to a site in the device system. We call the Hilbert space of the sites mapped to by $L$ the layout subspace of the device Hilbert space. A layout $L$ induces a mapping $\L$ from the target system's Hilbert space to the layout subspace. When limiting $\L(H)$ in the layout subspace where $H$ is a Hermitian matrix in the target Hilbert space, one can relabel the sites of $\L(H)$ according to $L^{-1}$ and recover $H$. When $H(t)$ is a time-dependent Hamiltonian of the target Hilbert space, we write $\L(H)$ as a Hamiltonian of the device Hilbert space satisfying $\L(H)(t)=\L(H(t))$. Let the execution of the instruction schedule $(\S, \vec{g})$ produce a unitary matrix $U$ and let the evolution under $\L(H_{\text{tar}})$ for time interval $[0, T)$ be $\L(U_{\text{tar}}).$ We say that a site layout $L$ and the instruction schedule $(\S, \vec{g})$ simulate $H_{\text{tar}(t)}$ if $U$ approximates $\L(U_{\text{tar}}).$

\subsubsection{Hamiltonian Synthesizer}

Since HML discretizes continuous Hamiltonians with small errors, in this step, we consider a target quantum system described by a sequence of evolution under $H_{\mathrm{tar}, j}$ for time duration $\tau_j$ indexed by $j\in\{1, ..., N\}$. We want to synthesize an instruction schedule simulating the target quantum system on a device described by an AAIS $D=[\eta_1; ...; \eta_M; H_{\mathrm{sys}}]$. Our Hamiltonian synthesizer follows a three-step loop: (1) propose a site layout $L$; (2) build a coefficient equation system; (3) solve the mixed-binary equation system. If the solver does not find an approximate solution, we repeat this process until a timeout condition is met.

\paragraph{Site layout proposer} The first step of the synthesizer loop proposes a site layout $L$ and later steps check its feasibility. To the best of our knowledge, although layout synthesis for quantum circuits is thoroughly studied \cite{tan2020optimal}, there is no prior work on the layout synthesis for Hamiltonian-oriented quantum computing. The main difference between them is the unavailability of swap gates for many analog quantum devices, i.e., QuEra's Rydberg atom arrays. 

We employ a search with pruning as a general solution to a layout proposer.
The pruning strategy is to abort the search when there exists a product Hamiltonian $P$ and $j$ where $H_{\mathrm{tar}, j}[P]\neq 0$ and $\L(P)$ does not have a non-zero coefficient expression in any $\eta_{k}$ and $H_{\mathrm{sys}}$. This abort condition can be met halfway through the search. For a partial layout $L$ (where several sites are not assigned in $L$ yet) and a product Hamiltonian $P$, we can map it to a product Hamiltonian $\L(P)$ of the device with holes on several sites. When searching for $\L(P)$ in an AAIS, holes can match any site operator. If none is found, the current search branch is aborted.

After proposing a layout, we proceed to steps (2) and (3) to check its feasibility. If rejected, the above search process returns and proceeds to other search branches to propose another layout. If all possibilities are not feasible, the compiler will report no solution and fail the process.

\begin{figure}
\scalebox{0.85}{
\begin{minipage}{0.52\linewidth}
\begin{algorithm}[H]
\caption{Equation builder for Hamiltonian synthesis.}
\label{alg:align}
\raggedright
\vspace{1pt}
\textbf{Inputs:} site layout mapping $\L$, target quantum system $(H_{\mathrm{tar}, j}, \tau_j)$ for $1\leq j\leq N$, AAIS $D=[\eta_1, ..., \eta_M, H_{\mathrm{sys}}]$, equation system variables $\{\vec{a}_{k,j}\}, \{s_{k,j}\}, \vec{g}, \{t_j\}$\\
\textbf{Output:} a system of equations $\Upsilon$
\vspace{2pt}
\hrule
\begin{algorithmic}
\State $\Upsilon\leftarrow\{\}$
\For{$j\in\{1, ..., N\}$}
    \State $G\leftarrow\{\L(H_{\mathrm{tar}, j})\}_{j=1}^N\cup\{H_{\mathrm{sys}}\}$ 
    \State $Q\leftarrow[P~|~\exists H\in G, H[P]\neq 0]$
    \State $i\leftarrow0$
    \While{$i<|Q|$}
        \State $P\leftarrow Q.\texttt{getitem}(i)$
        \State $i\leftarrow i+1$
        \If{$P=I$}
            \State \textbf{continue}
        \EndIf
        \State $e\leftarrow H_{\mathrm{tar}, j}[P](\vec{g})$ 
        \For{$k\in\{1,...M\}$}
            \If{$\ssem{\eta_k}[P]\not\equiv0$}
                \State $e\leftarrow e+\ssem{\eta_k}[P](\vec{a}_{k,j})\cdot s_{k,j}$
                \For{$P'\not\in Q~:~\ssem{\eta_k}[P']\not\equiv0$}
                    \State $Q.\texttt{append}(P')$
                \EndFor
            \EndIf
        \EndFor
    \State $\Upsilon.\texttt{add}(t_j\cdot e=\tau_j\cdot \L(H_{\mathrm{tar}, j})[P])$
    \EndWhile
\EndFor
\end{algorithmic}
\end{algorithm}
\end{minipage}
}
\quad
\begin{minipage}{0.47\linewidth}
    \includegraphics[width=\linewidth, trim=2cm 9cm 21cm 2cm, clip]{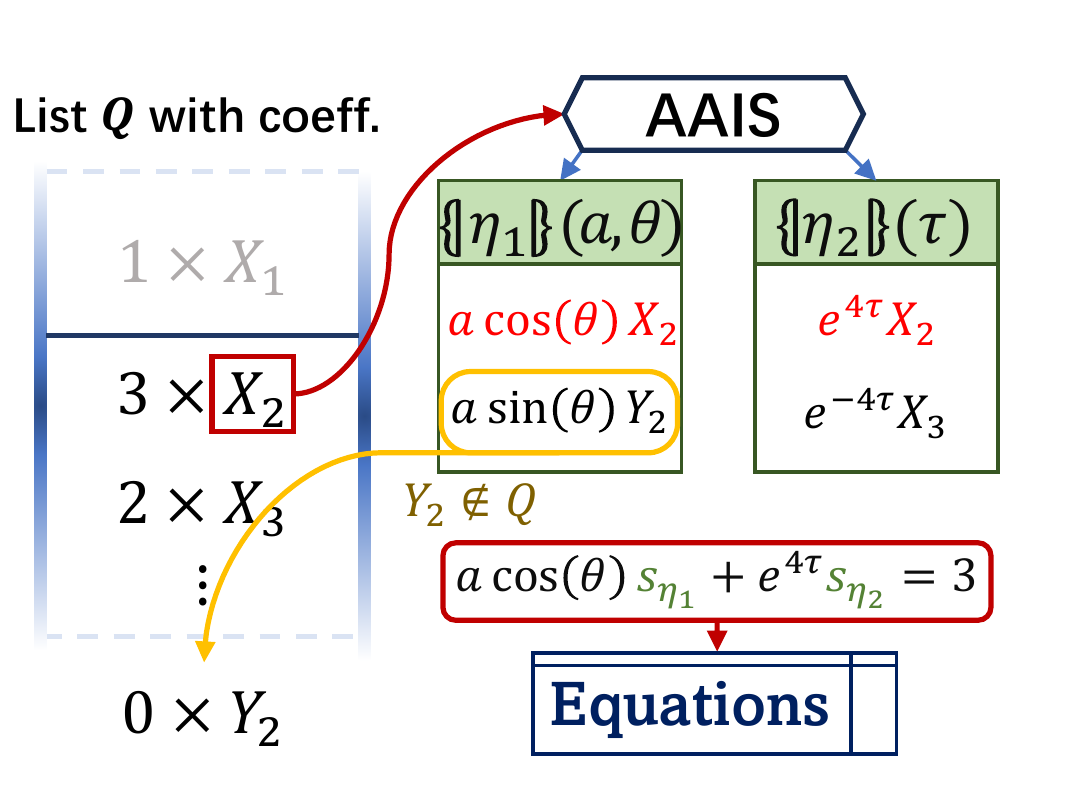}
    \vspace{-2em}
    \caption{An example illustrating the equation builder. $X_2$ is searched for in AAIS, where $\eta_1$ and $\eta_2$ are found and used in equation for $X_2$. }
    \label{fig:eqb}
    \par
    \includegraphics[width=\linewidth, trim=10cm 11cm 11cm 3.5cm, clip]{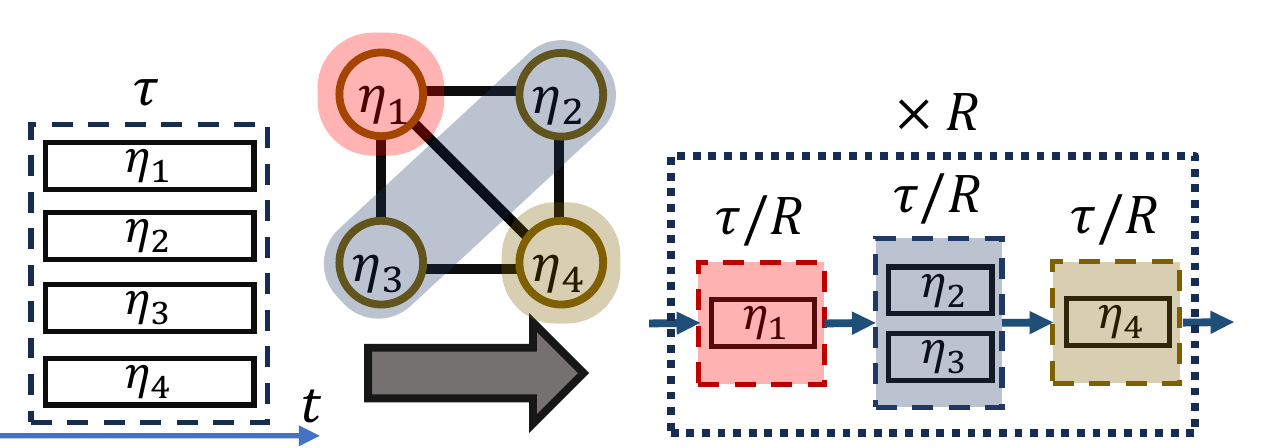}
    \vspace{-2em}
    \caption{An example of Trotterization from an instruction schedule to a block schedule, with a conflict graph and its grouping.}
    \label{fig:conflict}
\end{minipage}
\end{figure}

\paragraph{Coefficient equation builder} We synthesize instruction executions by a system of mixed-binary non-linear equations to match coefficients of product Hamiltonian in the target quantum system.

Given a site layout $L$, a set of equations is constructed to match the coefficients in $H_{\mathrm{tar}, j}$ for every $1\leq j\leq N$. We create time variables $t_j$ to represent the evolution time for instruction executions synthesizing evolution of $H_{\mathrm{tar}, j}$ for time $\tau_j$, with constraints $t_j>0$. For instruction $\eta_k$ in the AAIS, we create an indicator variable $s_{k,j}\in\{0, 1\}$ to indicate whether $\eta_k$ is selected to be executed in the synthesis of $H_{\mathrm{tar}, j}$. Assuming that $\eta_k$ has local variables $\vec{v}_k$ of dimension $|\vec{v}_k|$, we create $|\vec{v}_k|$ new equation system variables stored in a vector $\vec{a}_{k, j}$. 
For global variables, we create a vector $\vec{g}$ of dimension $|\vec{v}_{\mathrm{glob}}|$ of the AAIS, which is independent of $j$. In total, we have created $|\vec{v}_{\mathrm{glob}}|+N \sum_{k}|\vec{v}_k|+N$ real variables for global variables, local variables, and time variables respectively, and $NM$ indicator variables. 

Then we establish a coefficient equation for each product Hamiltonian $P$ to match $H_{\mathrm{tar}, j}$:
\begin{align}
    (\forall j), (\forall P\neq I): \quad t_j\cdot H_{\mathrm{sys}}(\vec{g})[P]+\sum\nolimits_{k=1}^{M}t_j\cdot\ssem{\eta_k}[P](\vec{v}_{k,j})\cdot s_{k,j}=\tau_j\cdot \L(H_{\mathrm{tar}, j})[P].
\end{align}
Here the left-hand-side calculates the summed effects of $P$ from each instruction and the system Hamiltonian and the right-hand-side calculates the effect of $P$ in the target quantum system.

There are typically many trivial equations in this system having $0$ on both sides since only a few $P$ appear in either $H_{\mathrm{tar}, j}$ or $\ssem{\eta_k}$ with respect to the exponentially many possible combinations of site operators. We propose \alg{align} to find all non-trivial equations. This algorithm starts with a list $Q$ containing all the product Hamiltonians with non-zero coefficients in $H_{\mathrm{sys}}$ and $\L(H_{\mathrm{tar}, j})$. It then enumerates the list $Q$ and establishes coefficient equations for each $P$ by enumerating instructions $\eta_k$ in AAIS. During this process, it may encounter instruction Hamiltonians $\ssem{\eta_k}$ who contain product Hamiltonians $P'$ that never appears in $Q$. These product Hamiltonians may lead to non-trivial equations, so we add them to $Q$. An example of this procedure is illustrated in \fig{eqb}. 

\paragraph{Mixed equation solver} The established coefficient equation system is mixed-binary and non-linear. A solver is applied to obtain approximate solutions which correspond to instruction schedules.

We provide several options for the solver. The first is dReal \cite{gao2013dreal} based on $\delta$-complete decision procedures, which supports real variables, binary variables, and algebraic functions in HML and AAIS-SL. It performs well when the coefficient expressions are close to linear (the Heisenberg AAIS), while poorly when highly non-linear (the Rydberg AAIS). 

As another option, we construct a least-squares-based solver. This solver uses a relaxation-rounding scheme. We apply a continuous relaxation to loosen the value range of indicator variables from $s_{k,j}\in\{0, 1\}$ to $\hat{s}_{k,j}\in[0, 1]$, substitute $\hat{s}_{k,j}$ for $s_{k, j}$ in the equation system, and solve the equation system by least-squares methods via an implementation in SciPy \cite{virtanen2020scipy}. We then round the indicator variables $s_{k,j}$ according to the solution. The criterion sets $s_{k,j}$ to 1 if there is $\sum_{P}t_j\ssem{\eta_{k,j}}[P](\vec{v}_{k,j})\hat{s}_{k,j}>\delta$ for a pre-defined tolerance parameter $\delta$, and sets to 0 otherwise. This criterion evaluates how much error the solution will induce if we set $s_{k, j}$ to $0$. We then solve the equation system again to obtain a more precise solution.

The solver generates an approximate solution with error $e$, defined by
\begin{align}
    e=\sum\nolimits_{k,j,P} \lvert\tau_j\ssem{\eta_k}[P](\vec{v}_{k,j})-t_j\L(H_{\mathrm{tar}, j})[P]\rvert.
\end{align}
If $e<\epsilon$ where $\epsilon$ is a pre-defined tolerance, the solution is accepted. Otherwise, we return to step (1) to generate another layout and check feasibility. An accepted solution induces an instruction schedule $(\S=\{(C_j, t_j)\}, \vec{g})$ where $C_j=\{(\eta_k, \vec{a}_{k, j}): s_{k,j}=1\}.$

\subsubsection{Error Induced by Hamiltonian Synthesizer}

Now we bound the error in the evolution induced by the approximation in the equation solving of the Hamiltonian synthesizer since our solver generates approximate numerical solutions.
Let $\hat{U}(T)$ be the unitary of executing generated instruction schedule $(\S, \vec{g})$, and $\tilde{U}'(T)=\L(\tilde{U}(T))$ be the unitary of the evolution of the discretized target system after site layout mapping $\L$. We can conclude the error induced by the Hamiltonian synthesizer is bounded by tolerance $\epsilon$ in the equation solving and the proof is routine and omitted.

\begin{lemma}[\cite{nielsen2002quantum}]
\label{lem:ham-syn}
The error of evolution induced by equation solving is bounded by a constant $C_2>0$ and error bound $\epsilon$ with the following inequality:
\begin{align}
    \norm{\tilde{U}'(T)-\hat{U}(T)}\leq C_2 \epsilon.
\end{align}
\end{lemma}

\begin{remark}
In general, compiling a target system is computationally hard. Finding a site layout for machines with specific topology can be as hard as the sub-graph isomorphism problem, an NP-complete problem. Besides, since the design of AAIS does not pose strict restrictions on the expressions, pathological functions may emerge in the coefficients, which complicates the equation-solving process. Our solutions to these problems may not be optimal but are intuitive, feasible, and efficient enough for most cases (also refer to \sec{cs} for detailed case studies).
\end{remark}

\subsection{Block Schedules and Conflict Resolver}
\label{sec:conflict}

Instruction schedules are oversimplified descriptions of what can be executed on the devices. Mainly, there are two realistic restrictions not captured by instruction schedules. First, some instructions on real devices can not be executed simultaneously. For example, on an IonQ device, $\eta_{1, 2, XX}$ cannot be simultaneously executed with $\eta_{1, 2, ZZ}$ since they use the same interaction process with different bases. Second, instruction execution implementations may take longer than the scheduled execution time. We propose a flexible generalization to the instruction schedules called block schedules and implement a conflict resolver to compile generated instruction schedules to block schedules.

\subsubsection{Block Schedules}

A block schedule is a temporal graph whose vertices are blocks of instruction executions, together with the valuation of the global variables. An instruction block $B$ contains a collection of instruction executions whose evolution duration is $\tau$. The block schedule is then a directed acyclic graph where an edge $(B_j\rightarrow B_k)$ is a restriction: instructions in $B_k$ should start simultaneously after instruction executions in $B_j$ end. 
Instruction schedules generated by our Hamiltonian synthesizer are special cases of block schedules where the temporal graph forms a chain and blocks are the collections of instruction executions.

When executing a block schedule, we first decide the execution order $\gamma:(B_1, ..., B_r)$ of the blocks and then evolve the system by $\gamma$ sequentially. Let the $B_j$ contain $\{(\eta_{j,k}, \vec{a}_{j,k})\}_k$ with evolution time $\tau_j$. The evolution will generate a unitary transformation
\begin{align}
    U_{\gamma}=\prod\nolimits_{j=r}^1e^{-i\tau_j(H_{\mathrm{sys}}(\vec{g})+\sum_k\ssem{\eta_{j,k}}(\vec{a}_{j, k}))}.
\end{align}
Our next step is to generate a block schedule where instructions in each block are simultaneously executable and approximate the execution of the instruction schedule.

\subsubsection{Instruction Decorations}
\label{sec:decoration}

In general, the conflict relation of instructions forms a graph $F$: $(\eta_j, \eta_k)\in F$ means that $\eta_j$ and $\eta_k$ cannot be executed simultaneously. To ease the description of $F$, we introduce decorations to instructions to specify properties like categories of instructions. More decorations can be added based on the detailed hardware restrictions accordingly. 

\paragraph{Signal Lines}
Physical pulses are sent to devices through signal carriers like electronic wires or arbitrary waveform generators (AWG). A natural conflict is that if two instructions require the same signal carrier, they cannot be executed simultaneously. We abstract the concept of signal carriers as \emph{signal lines} and assign each instruction $\eta$ to a signal line denoted by $SL(\eta)$. If $SL(\eta_j)=SL(\eta_k),$ instructions $\eta_j, \eta_k$ conflict with each other.

\paragraph{Nativeness}
Another aspect leading to conflicts is whether instruction implementations employ compound pulses to approximate an effective Hamiltonian. For example, IBM devices generate $\ssem{\eta_{j, k, CR}}$ by direct microwave controls of a cross-resonance pulse \cite{malekakhlagh2020first}. Hence the IBM-native AAIS for IBM devices has $\eta_{j,k,CR}$ as \emph{native} instructions: they can be simultaneously executed with other native instructions. To effectively realize $\ssem{\eta_{j, k, ZZ}}$, a compound sequence of microwaves including two cross-resonance pulses is applied to approximate a $Z_jZ_k$ interaction \cite{alexander2020qiskit}. Simultaneously applying other pulses on site $j$ or $k$ will break the approximation. Hence $\eta_{j,k,ZZ}$ are \emph{derived} instructions in the IBM-native AAIS. 

Let $\inf{H}$ be the sites on which Hamiltonian $H$ acts non-trivially (when limited on these sites, $H$ is not identity). We assume that implementing a derived instruction $\eta$ only affects $\inf{\ssem{\eta}}$. Then a derived instruction $\eta_1$ conflicts with $\eta_2$ if $\inf{\ssem{\eta_1}}\cap \inf{\ssem{\eta_2}}$ is not empty.

\subsubsection{Conflict Resolver via Trotterization}

Given a conflict graph and an instruction schedule $(S, \vec{g})$, we implement a conflict resolver to generate a block schedule without conflicts in each block.

A well-studied technique in quantum information to simulate summed Hamiltonians in quantum simulation is Trotterization. Let Hamiltonian $H=\sum_{j=1}^L H_j$ where $L$ Hamiltonians evolve the system simultaneously. We assume we have a device supporting evolving single $H_j$ for any duration $t$, realizing unitary matrix $e^{-itH_j}$, while there is no evolution under $\sum_{j=1}^L H_j$. Trotterization (also known as the product formula algorithm) \cite{lloyd1996universal} makes use of the Lie-Trotter formula 
\begin{align}
\label{eq:trotter}
    e^{-it\sum_j H_j}=\lim\nolimits_{n\rightarrow\infty}\left(\prod\nolimits_j e^{-i\frac{t}{n}H_j}\right)^n\approx \left(\prod\nolimits_j e^{-i\frac{t}{N}H_j}\right)^N.
\end{align}
By choosing a large $N$, the above formula shows that we can approximate the evolution under $H$ for time $T$ by repeating for $N$ times a sequential evolution for $j\in\{1, ..., L\}$ under $H_j$ for time $t/N$. 
Each segment of evolution realizes a unitary transformation $e^{-i(t/N)H_j}$ as in the formula. 

First, we consider the case where $H_{\mathrm{sys}}=0$. Each $(C_d, \tau_d)$ in $\S$ is considered independently. Let $C_d=\{(\eta_j, \vec{a}_j)\}_j$ and the conflict graph of these instructions be $F$. To accommodate Trotterization in a conflict resolver, we first categorize the instructions into groups without conflict. The grouping is effectively a coloring of vertices in $F$ where no edge connects monochromatic vertices. We employ a greedy graph coloring algorithm from NetworkX \cite{hagberg2008exploring} to find a feasible grouping $\{G_j\}_{j=1}^L$ with $L$ colors where $G_j$ contains instruction executions in the $j$-th group. 

A temporal graph in a block schedule can depict the process in \eq{trotter}. Let $H_j$ be the Hamiltonian of simultaneous instruction executions in $G_j$, $H_j=\sum_{(\ssem{\eta_k}, \vec{a}_k)\in G_j} \ssem{\eta_k}(\vec{a}_k)$, and $R$ be the Trotterization number specified by users. An evolution of $H_j$ for time $\tau_d/R$ corresponds to a block $B_j=(G_j, \tau_d/R)$. Then a sequential evolution of $\{H_j\}_{j=1}^L$ forms a chain $B_1\rightarrow \cdots \rightarrow B_L$. We create $R$ copies of this chain and connect them sequentially to represent the Trotterization process.

Additionally, we deal with the cases where the system Hamiltonian $H_{\mathrm{sys}}$ is non-zero. Let $\tilde{L}$ be the maximal coloring number $L$ in the above process. We assume that there exists $\vec{g}_{\tilde{L}}$ such that $H_{\mathrm{sys}}(\vec{g}_{\tilde{L}})=H_{\mathrm{sys}}(\vec{g})/\tilde{L}.$
Some devices may not support this assumption, but it is rarely used since only a few devices with non-zero system Hamiltonian have conflicting instructions. We then augment the number of groups to $\tilde{L}$ for each $(C_j, \tau_j)\in\S$ by adding empty sets in groupings. Now we create a block schedule with $\vec{g}_{\tilde{L}}$ and a temporal graph constructed on the augmented groupings. Executing this block schedule approximates the execution of the given instruction schedule.

\subsubsection{Error Induced by Conflict Resolver}

The Trotterization resolves conflicts while also introducing errors. We denote the instruction schedule where $S=\{(\{(\eta_{d, j}, \vec{a}_{d, j})\}_j, \tau_d)\}_{d=1}^D$ and its evolution as $\hat{U}(T)$. For segment $d$ of evolution in $S$, we assume the grouping is $\{G_j^d\}_{j=1}^{L_d}$ and the evolution by executing the block schedule as $\bar{U}(T)$.

\begin{lemma}[\cite{pnas-simulation}]
\label{lem:trotter}
The difference between $\hat{U}(T)$ and $\bar{U}(T)$ of evolution after resolving conflicts by Trotterization is bounded by
\begin{align}
    \norm{\hat{U}(T)-\bar{U}(T)}\leq \frac{(\Lambda T)^2}{DR}e^{\frac{\Lambda T}{DR}}.
\end{align}
Here $\Lambda=\max_{d, j}L_d\norm{\sum_{(\eta, \vec{a})\in G_j^d}\ssem{\eta}(\vec{a})}$, $D$ and $R$ are the discretization and Trotterization numbers.
\end{lemma}

As implied by this lemma, in ideal cases, increasing the Trotterization number reduces the induced error to arbitrarily small. However, it also increases the total number of instruction executions. Due to the non-negligible error accumulations in each instruction execution on devices, there is a trade-off over the Trotterization number $R$ depending on the real-time parameters of the device, where we leave the freedom to user specification.

Optimization techniques for Trotterization are also well-developed in theory \cite{childs2021theory}, and we discuss their implementation in \app{trotter-opt}.

\subsection{Signal Line Schedules and Scheduler}

The eventual output of SimuQ contains the pulses sent through signal carriers for devices to execute. We propose another intermediate representation, called a \emph{signal line schedule}, to depict the concrete instruction executions sent through each signal line abstracted in \sec{decoration} before generating platform-dependent executable pulses. For signal line $l$, it contains a list of instruction executions $(\eta, \vec{a}, \tau_s, \tau_e)$ with absolute starting and ending times and satisfying $SL(\eta)=l$. 

To obtain a signal line schedule, we build a scheduler to traverse the temporal graph of the block schedule via a topological sort and generate a valid execution order of block schedules. It employs a first-arrive-first-serve principle for each signal line. The scheduler first extracts information about how long implementing each instruction execution takes from real devices. Next, it arranges instruction executions on the signal lines at the earliest possible starting time obeying the order. 

\begin{remark}
The scheduling process may be independently configured and optimized, and the scheduler may use other criteria to determine the traversal order of instruction blocks or the alignment of blocks within the scheduled order as long as the hardware permits. This freedom in the scheduling process may be leveraged to reduce cross-talk \cite{murali2020software} between the blocks or save small implementation overheads. We illustrate only a basic strategy and leave the exploitation for the future.
\end{remark}

\subsection{Pulse Schedules and Pulse Translator}

In its final stage, the SimuQ compiler translates a signal line schedule into a pulse schedule using hardware providers' domain languages and APIs. 

We extract the pulse shapes from the devices for each platform to implement instruction executions. We substitute the instruction execution on each signal line for pulse shapes configured by the valuations of local variables via the format specified by a pulse-enabled quantum device provider.

\subsubsection{Translation to Hardware APIs}

There are few pulse-enabled quantum device providers, and programming pulses is a challenging endeavor that requires extensive platform knowledge of various hardware and software engineering considerations. We demonstrate the effectiveness of SimuQ using QuEra, IBM, and IonQ devices. 

\paragraph{QuEra's Rydberg atom devices} Two APIs to QuEra devices are supported by SimuQ for the global Rydberg AAIS: Bloqade \cite{Bloqade} programs and Amazon Braket programs. We set the atom positions according to the valuation of global variables and laser configurations as piecewise constant functions according to the valuations of local variables. Since the detuning $\Delta$ and amplitude $\Omega$ generate linear effects, piecewise linear laser configurations are also supported as an option. For Amazon Braket programs, $\Omega(t)$ should start and end at amplitude $0$, so we add short (0.1ms) time intervals to the pulses' beginning and end with linear ramping. We also scale the pulse schedules to a total length of around $3.5$ms to fit in the $4$ms duration limit of the device.

\paragraph{IBM's superconducting devices} For IBM devices, SimuQ can generate Qiskit Pulse programs for the Heisenberg AAIS and the IBM-native AAIS. For single-site instructions, the IBM device supports implementations of native $X$ and $Y$ instructions and derived $Z$ instructions. We build up DRAG pulses \cite{motzoi2009simple} to realize $X$ and $Y$ instructions and free Z rotations \cite{mckay2017efficient} to realize $Z$ instructions, which are standard superconducting device techniques. Two-qubit instructions in the Heisenberg AAIS are realized through the $Z_jX_k$ interactions created by echoed cross resonance pulses \cite{malekakhlagh2020first} together with single-qubit evolution to change bases. We follow \citet{earnest2021pulse} and realize interaction-based gate implementations, whose benefits are further explained in \sec{cs-qaoa}. Additionally, we extract cross-resonance pulses from Qiskit and compose pulses to realize native $\eta_{j, k, CR}$ in the IBM-native AAIS.

\paragraph{IonQ's trapped-ion devices} SimuQ supports both IonQ cloud and Qiskit circuit programs for IonQ devices with the Heisenberg AAIS. Unlike QuEra and IBM devices, IonQ does not provide pulse-level programmability for their ion trap devices. However, we can still exploit their native gate set to generate a quantum circuit with precise control of the execution on their devices. With the support of partially entangling M\o lmer-S\o renson gate \cite{sorensen2000entanglement}, we can implement instructions of the Heisenberg AAIS with higher fidelity. More details are explained in our case studies in \sec{cs-qaoa}. 

\subsubsection{Semantics of Pulse Schedules and Errors in Instruction Implementation}

Abstractly, a pulse schedule includes a time-dependent function $\vec{f}_l(t)$ (pulses) for signal line $l$, generating the effective Hamiltonian $H_l(t)$ physically. For example, instruction execution $(\eta, \vec{a}, \tau_s, \tau_e)$ for signal line $l$ in the signal line schedule should be translated into pulses $\vec{f}_l(t)$ that effectively generate $H_l(t)=\ssem{\eta}(\vec{a})$ for $\tau_s\leq t<\tau_e$. Collectively, the Hamiltonian on the device is $H_{\mathrm{dev}}(t)=H_{\mathrm{sys}}+\sum_l H_l(t),$ and the semantics of executing a pulse schedule is the unitary evolution under $H_{\mathrm{dev}}(t)$. Yet, the implementation of instructions on real devices may be imperfect. We assume that there is a implementation error threshold $\Delta$ such that the on-device $\check{H}_l(t)$ and $\check{H}_{\mathrm{sys}}$ satisfies $\max_{t, l}\norm{\check{H}_l(t)-H_l(t)}\leq \Delta$ and $\norm{\check{H}_{\mathrm{sys}}-H_{\mathrm{sys}}}\leq \Delta$, forming on-device evolution under $\check{H}_{\mathrm{dev}}(t)=\check{H}_{\mathrm{sys}}+\sum_l\check{H}_l(t).$ Since the signal line scheduler does not alter the semantics of block schedules, we bound the implementation error on the device.

\begin{lemma}[\cite{nielsen2002quantum}]
\label{lem:implement}
    The difference between the unitary $\check{U}(T)$ on the device and the unitary $\bar{U}(T)$ of executing the block schedule generated by the conflict resolver is bounded by
    \begin{align}
        \norm{\bar{U}(T)-\check{U}(T)}\leq C_3S\Delta \Gamma T,
    \end{align}
    where $C_3>0$ is a constant, $S$ is the number of signal lines and system Hamiltonians, and $\Gamma=\max_d L_d$ is the maximal number of groups in the conflict resolver.
\end{lemma}

With a faithful implementation of instructions on real devices, the pulse translator produces negligible errors. The proof is routine in quantum information and is therefore omitted. We remark that other forms of device errors (e.g., high-energy space leakage) can be analyzed similarly.

\subsection{Semantics Preservation of SimuQ Compiler}

If compilation succeeds, the SimuQ compiler generates executable pulse schedules from programmed quantum systems with bounded errors. We conclude the approximate semantics preservation theorem of the SimuQ compilation process using \lem{discretize}, \lem{ham-syn}, \lem{trotter}, and \lem{implement}.

\begin{theorem}[Semantics Preservation]
Given a Hamiltonian $H_{\mathrm{tar}}(t)=\sum_{k=1}^K\alpha_k(t)H_k$ where $\alpha_k$ is piecewise $M$-Lipschitz and $\norm{H_k}=1$, if the compilation succeeds, SimuQ generates a site layout $L$ and an executable pulse schedule. Let the unitary $U(T)$ represent the evolution under $H_{\mathrm{tar}}(t)$ for duration $[0, T]$ and $\check{U}(T)$ for the evolution executing the pulse schedule on the device. We have
\begin{align}
    \norm{\L(U(T))-\check{U}(T)}\leq C_1D^{-1}MKT^2+C_2\epsilon+(\Lambda T)^2D^{-1}R^{-1}e^{\frac{\Lambda T}{DR}}+C_3S\Delta \Gamma T.
\end{align}
Here, $T$ is the evolution time, $\L$ is the site layout mapping of layout $L$, $C_1, C_2, C_3$ are constants, $D$ is the discretization number, $\epsilon$ is the error threshold in Hamiltonian synthesizer, $R$ is the Trotterization number, $\Delta$ is the instruction implementation error threshold, $S$ is the number of signal lines and system Hamiltonians on the device, and $\Lambda$ and $\Gamma$ depend on the Trotterization strategy in the compilation.
\end{theorem}

Tuning $D, \epsilon, R$ and improving the implementation to decrease $\delta$ can reduce errors induced by the SimuQ compiler to arbitrarily small. The evolution under $H_{\mathrm{tar}}$ is hence simulated on the device.

%% file: 5-CaseStudy.tex
\section{Case Studies} \label{sec:cs}

We conduct several case studies highlighting SimuQ's portability and the advantages of Hamiltonian-oriented compilation, including native instructions and interaction-based gates. We also establish a small benchmark of quantum simulation to evaluate the SimuQ compiler performance. 

\subsection{Multiple-Platform Compatability}
\label{sec:cs-ising}

We compile and execute the Ising model on multiple supported devices of SimuQ. The following experiments show the portability of SimuQ on heterogeneous analog quantum simulators. We only need to program the target quantum systems once and apply the SimuQ compiler to generate code for different platforms and deploy and execute them on multiple real devices.

We focus on the simulation of the Ising model introduced in \sec{exp-ising}. We demonstrate two instances: a 6-site cycle and a 6-site chain, mathematically depicted by
\begin{align}
    H_{\mathrm{chain}}=\sum\nolimits_{j=1}^5 Z_jZ_{j+1}+\sum\nolimits_{j=1}^6X_j, \quad H_{\mathrm{cycle}}=H_{\mathrm{chain}}+Z_1Z_6.
\end{align}
The target quantum system is to simulate $H_{\mathrm{cycle}}$ and $H_{\mathrm{chain}}$ for $T=1$. When Trotterization is utilized, we set the Trotterization number to be $4$, which is empirically selected based on experiment results. 

\begin{figure}[t]
    \centering
    \includegraphics[scale=0.7, trim=1cm 12.5cm 11cm 2.1cm, clip]{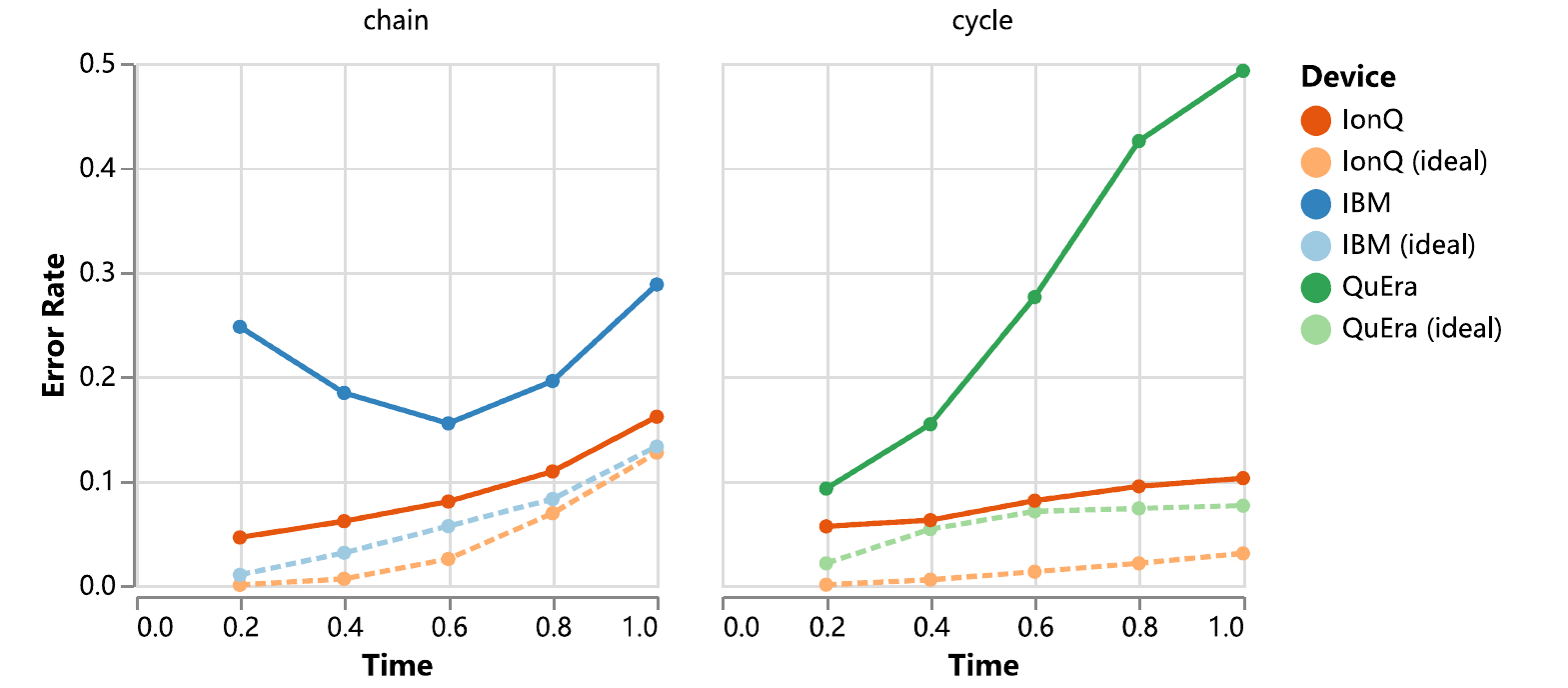}
    \vspace{-1.2em}
    \caption{The simulation errors of the 6-site Ising models on multiple platforms. Ideal results are obtained by compiling with SimuQ and executing on noiseless simulators.}
    \vspace{-1em}
    \label{fig:ising-sim}
\end{figure}

SimuQ successfully compiles $H_{\mathrm{cycle}}$ on QuEra devices using the global Rydberg AAIS, both $H_{\mathrm{cycle}}$ and $H_{\mathrm{chain}}$ on IonQ devices using the Heisenberg AAIS, and $H_{\mathrm{chain}}$ on IBM devices using the Heisenberg AAIS. We send the generated code to execute on corresponding devices. Since QuEra devices do not support state tomography, we evaluate the results on these platforms by a metric based on measurements supported by all devices in our experiments. We obtain the frequency of reading a bit-string $s$ in a measurement instantly after the simulation finishes as a distribution $\mathbb{P}_{\mathrm{exp}}[s]$ and numerically calculate the ground truth distribution $\mathbb{P}_{\mathrm{GT}}[s]$ of obtaining $s$. We utilize the total variation distance $TV(\mathbb{P}_{\mathrm{exp}}, \mathbb{P}_{\mathrm{GT}})=\frac{1}{2}\sum\nolimits_{s\in\{0, 1\}^6}\lvert\mathbb{P}_{\mathrm{exp}}[s]-\mathbb{P}_{\mathrm{GT}}[s]\rvert$ to evaluates the errors.
We present the classical simulation of devices and real device execution results in \fig{ising-sim}. 

The minor errors in the classical simulations indicate the correctness of our framework. In ideal cases, the errors are induced by uncancellable non-neighboring $Z_jZ_k$ interactions and short ramping times for the global Rydberg AAIS and by Trotterization errors for the Heisenberg AAIS. The real device execution results show higher errors than classical simulation because of device noises, while they are valid quantum simulation results. The errors on the IBM device are non-monotone, likely because the large state preparation and measurement errors affect more heavily the cases where the states deviate only a little from the initial state.

SimuQ fails to compile $H_{\mathrm{chain}}$ on QuEra devices since it requires different local detuning parameters for different sites, which current QuEra devices and the global Rydberg AAIS do not support. It also fails to compile $H_{\mathrm{cycle}}$ on IBM devices since there is no 6-vertex cycle in IBM devices.

\subsection{Hamiltonian-Oriented Compilation with Native Instructions}
\label{sec:cs-native}

The most significant benefit of enabling Hamiltonian-level programming is to gain fine-grained and multi-site control via native operations. Near-term quantum devices have short coherence times: quantum states will deteriorate and lose their quantumness quickly. Generating shorter pulses to achieve the same effects is one of the crucial tasks for compilers of modern quantum devices. In this case study, we showcase the advantage in the lengths of pulse schedules enabled by Hamiltonian-oriented compilation using the IBM-native AAIS.

Our target quantum system evolves under $H_{2ZX}=Z_1X_2+X_2Z_3$ for time $T=1$, a small 3-site system. 
The IBM-native AAIS contains two native instructions $\eta_{1, 2, CR}$ and $\eta_{3, 2, CR}$ with $Z_1X_2$ and $X_2Z_3$ interactions respectively. Following \citet{greenaway2022analogue}, the simultaneous execution of them can be realized by simultaneously applying two cross-resonance pulses on IBM devices. By automatically compensating the other terms in SimuQ with native instructions $\eta_{2, X}$ and derived instructions $\eta_{1, Z}$ and $\eta_{2, Z}$ (their effects commute with $\ssem{\eta_{1, 2, CR}}$ and $\eta_{3, 2, CR}$ so no Trotterization is needed), the uncancellable remains are $Z_1Z_2$ interactions and $Z_2Z_3$ interactions. Fortunately, they can be reduced to one magnitude smaller than $Z_1X_2$ and $X_2Z_3$ interactions when selecting a relatively large $
\Omega$, and are considered small errors in the compilation. The pulse schedule to realize $H_{2ZX}$, displayed in \fig{Hzx-simuq}, is around $280$ns long.

$H_{2ZX}$ can also be compiled on IBM devices by a circuit-based compilation with the help of Qiskit. It first decomposes the simulation into a circuit with two gates $R_{Z_1X_2}(2)R_{X_2Z_3}(2)$ where $R_{Z_jX_k}(\theta)=e^{-i(\theta/2)Z_jX_k}$. It then invokes Qiskit's transpiler to decompose each $R_{Z_jX_k}(\theta)$ into two CNOT gates and several single qubit gates and generates a Qiskit pulse schedule, which is displayed in \fig{Hzx-qiskit} and is around $1660$ns long. This is around six times longer than the pulse schedule generated by SimuQ using the IBM-native AAIS.

\input{tabs/qaoa_result}

\subsection{Hamiltonian-Oriented Compilation with Interaction-based Gates}
\label{sec:cs-qaoa}

For some devices that lack the support of simultaneous instruction executions by native operations, we can still exploit the capability of realizing gates based on evolving interaction for various time periods. By interaction-based gates, we refer to quantum gates of form $R_H(t)=e^{-itH}$, where the time duration of the pulse shapes implementing them is strongly correlated with $t$. These gates are common on platforms supporting universal gates like IBM devices and IonQ devices but are not exploited in their provided compiler due to the hardness in calibration. 
Under the conventional circuit-oriented compilation where quantum programs are compiled to a gate set with fixed number 2-qubit gates (typically, only CNOT gates), interaction-based gates are decomposed using multiple 2-qubit gates for the convenience of calibration, like $R_{Z_jX_k}(\theta)$ gates that are decomposed using 2 CNOT gates by Qiskit and translated to a pulse schedule of long and fixed duration.
Although interaction-based gates cannot be simultaneously applied on devices when overlapping sites exist, exploiting them can still significantly reduce the duration of generated pulse schedules and increase the fidelity of simulations as observed in \cite{earnest2021pulse, stenger2021simulating}.

In this section, we implement the quantum approximate optimization algorithm (QAOA) in SimuQ and execute it on IBM and IonQ devices. The QAOA algorithm is a classical-quantum Hamiltonian-oriented algorithm designed to solve combinatorial problems. We omit the algorithm analysis and refer interested readers to \cite{farhi2014quantum}. 
We consider the quantum simulation part of a typical case of the QAOA algorithm, where the target quantum system evolves under a length-$p$ sequence of alternative evolution between $H_1$ and $H_2$ where
\begin{align}
    H_1=Z_1Z_N+\sum\nolimits_{j=1}^{N-1} Z_j Z_{j+1}, \qquad H_2=\sum\nolimits_{j=1}^N X_j.
\end{align}
Here $N=12$ is the problem size. Two pre-defined parameter lists $\{\theta_j\}_{j=1}^p$ and $\{\gamma_j\}_{j=1}^p$ of length $p$ describe the time of each evolution segment. I.e., the $j$-th segment first lets the system evolve under $H_1$ for time $\theta_j$ and then lets the system evolve under $H_2$ for time $\gamma_j$. Ultimately, we measure the sites and store the results in a bit-string $s$. The more precisely we simulate the system, the larger the evaluation function $C(s)=|\{j: 1\leq j\leq N, s_j\neq s_{j+1}\}|$ will be (assuming $s_{N+1}=s_1$).

When compiling the target system for the Heisenberg AAIS, $\eta_{j, j+1, ZZ}$ are executed frequently. We take an execution $\eta_{1, 2, ZZ}$ for $t=1$ as an example, which effectively realizes the gate $e^{-iZ_1Z_2}$. Qiskit decomposes it with 2 CNOT gates and single qubit rotation gates, and the generated pulse duration takes a constant $662$ns independent of $t$, as illustrated in \fig{pulse-qiskit}. An alternative solution is to create $Z_1X_2$ interaction constructed by echoed cross-resonance pulses for a duration positively correlated to $t$ with short pulses implementing Hadamard gates to effectively realize the $Z_1Z_2$ interaction, as illustrated in \fig{pulse-simuq}. The pulse schedule is around $(200t+130)$ns long to execute $\ssem{\eta_{j, j_1, ZZ}}(1)$ for time $t$. When $t=1$, it is $324$ns long, $51\%$ shorter than the Qiskit compiler. Interaction-based gates are especially beneficial when a program requires many short instruction executions, like compiled simulations with a large Trotterization number.

We then compile and execute the QAOA system simulation to IBM and IonQ devices for cases $p=1, 2, 3$. Similarly, we reproduce this problem in Qiskit and compile it with the Qiskit compiler (CNOT-based decomposition is applied). On IBM devices, for $p=3$, the pulse schedule generated by SimuQ is $3.35$ms long. In contrast, the one generated by Qiskit is $7.48$ms long, which is more than two times longer\footnote{IonQ devices do not support reporting pulse schedule duration.}. On average, SimuQ generates pulse schedules 59\% percent shorter than Qiskit.
We then execute the generated programs on IBM devices and IonQ devices. The differences of the evaluation function $C(s)$ measured on devices and ground truth values are present in \tab{measured_cut}. We observe that, on average, the pulse schedules generated by SimuQ reduce errors (the difference to ideal results) by 34\% on IBM devices and 28\% on IonQ devices for $p=3$ compared to pulse schedules from Qiskit. The advantage is less significant for shallow cases where $p=1, 2$ because of state preparation and measurement errors on real devices \cite{tannu2019mitigating}. These experiments demonstrate the advantage and the necessity of Hamiltonian-oriented compilation using interaction-based gates on devices not supporting simultaneous instruction executions.

\subsection{Benchmarking Quantum Simulation Compilation}
\label{sec:cs-benchmark}

\input{tabs/benchmark}

To illustrate SimuQ's capability of dealing with various quantum simulation problems, we craft a small benchmark containing models collected from multiple domains like condensed matter physics, high-energy physics, particle physics, and optimization. The diversity of the cases in this benchmark of different topologies, time dependency, and system sizes exhibits our compiler's feasibility and efficiency in dealing with significant simulation problems. 

We present the benchmark in \tab{benchmark}, and further illustrations of its quantum systems are in \app{benchmark}. We report each quantum system's number of sites and lines of code to implement them with HML in SimuQ. Most systems can be programmed within 20 lines, showing the user-friendliness of programming quantum systems in SimuQ. 

For each target quantum system, we compile it on the platforms supported by SimuQ using their most capable devices in the possible future. The compilation time is averaged over 5 runs on a laptop with Intel Core i7-8705G CPU. SimuQ compiler reports no solution (No sol.) in several cases due to complicated interactions beyond the hardware capability of QuEra devices and the limited connectivity of IBM devices. Limited connectivity on large IBM devices also complicates the site layout search, making a case exceed a pre-set compilation time limit of 3600 seconds, which is marked as a time out.

Pulse schedule duration for IBM devices using SimuQ and Qiskit to compile is reported. On average, Qiskit's default compilation passes generate 29.3 times longer pulse schedules than the SimuQ compiler over cases successfully compiled. We also report the number of partially entangling M\o lmer-S\o renson gates when compiling on IonQ's devices to indicate the total duration.

%% file: tabs/qaoa_result.tex
\begin{table}
\quad
\scalebox{0.9}{
\begin{minipage}{.45\linewidth}
\centering
    \includegraphics[width=\linewidth]{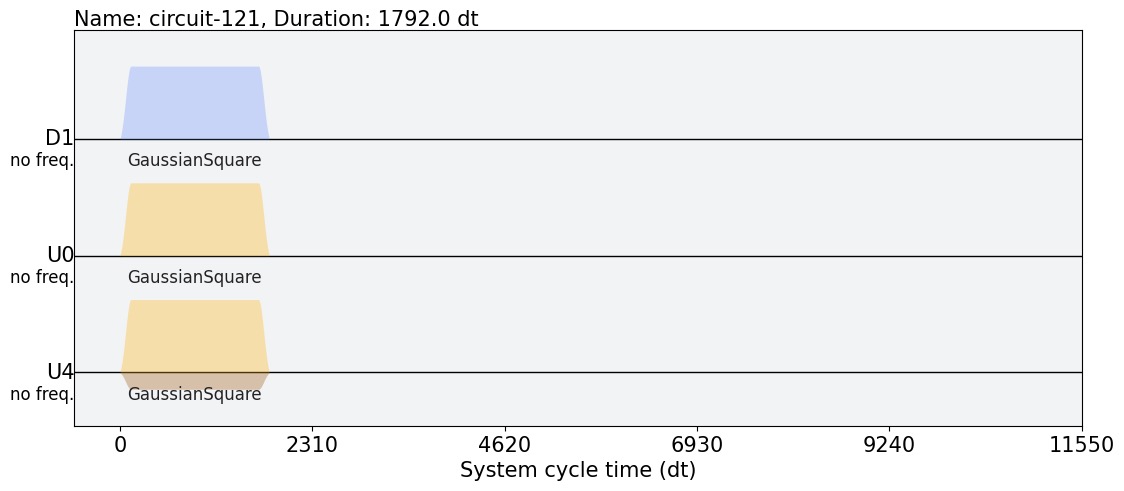}
    \vspace{-1em}
    \captionof{figure}{Pulses generated by SimuQ for evolution under $H_{2ZX}$ for duration $1$.}
    \label{fig:Hzx-simuq}
    \includegraphics[width=\linewidth]{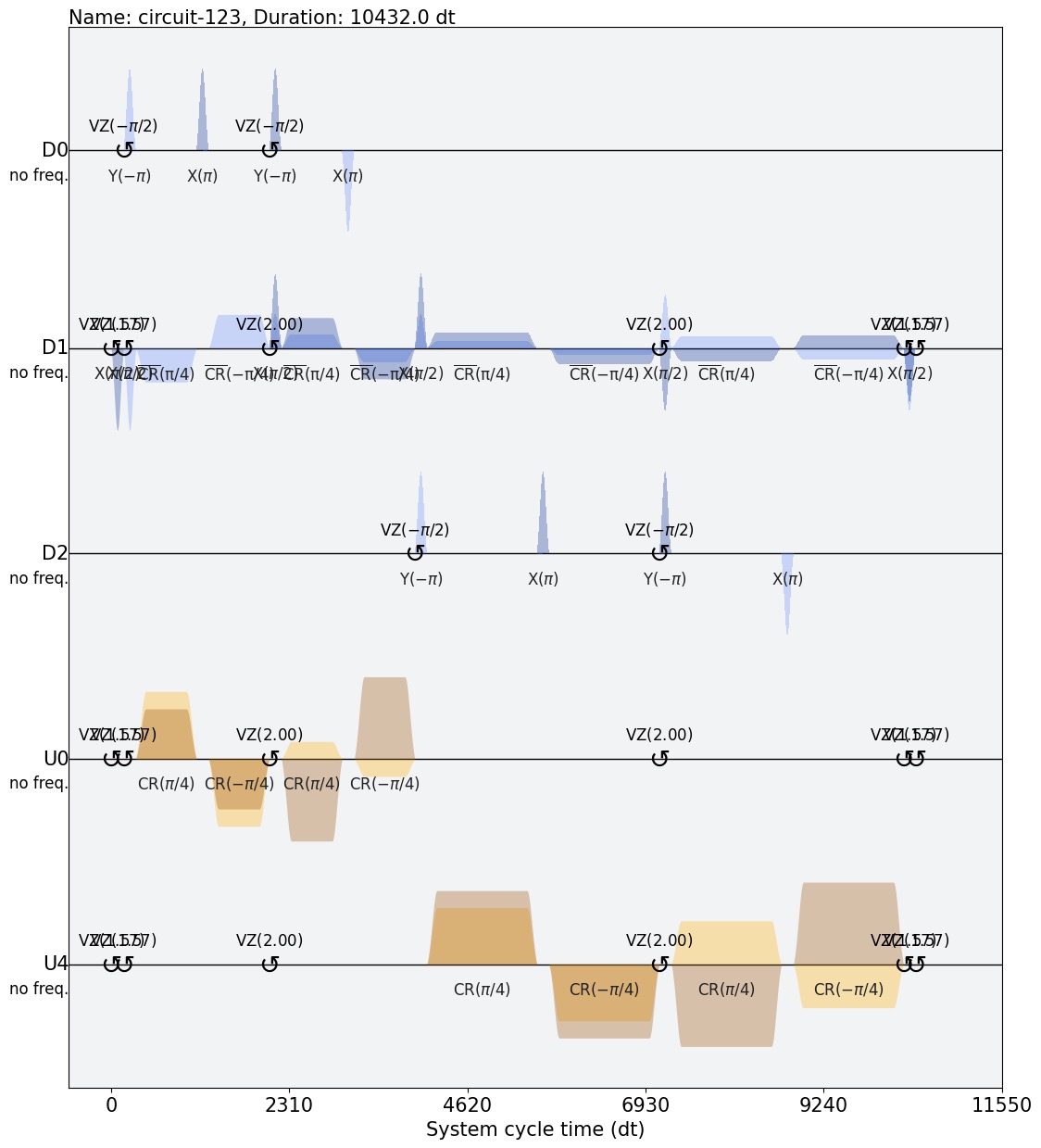}
    \vspace{-1em}
    \captionof{figure}{Pulses generated by Qiskit compiler for evolution under $H_{2ZX}$ for duration $1$.}
    \label{fig:Hzx-qiskit}
\end{minipage}
}
\hfill
\scalebox{0.9}{
\begin{minipage}{.6\linewidth}
\centering
    \includegraphics[width=0.70\linewidth]{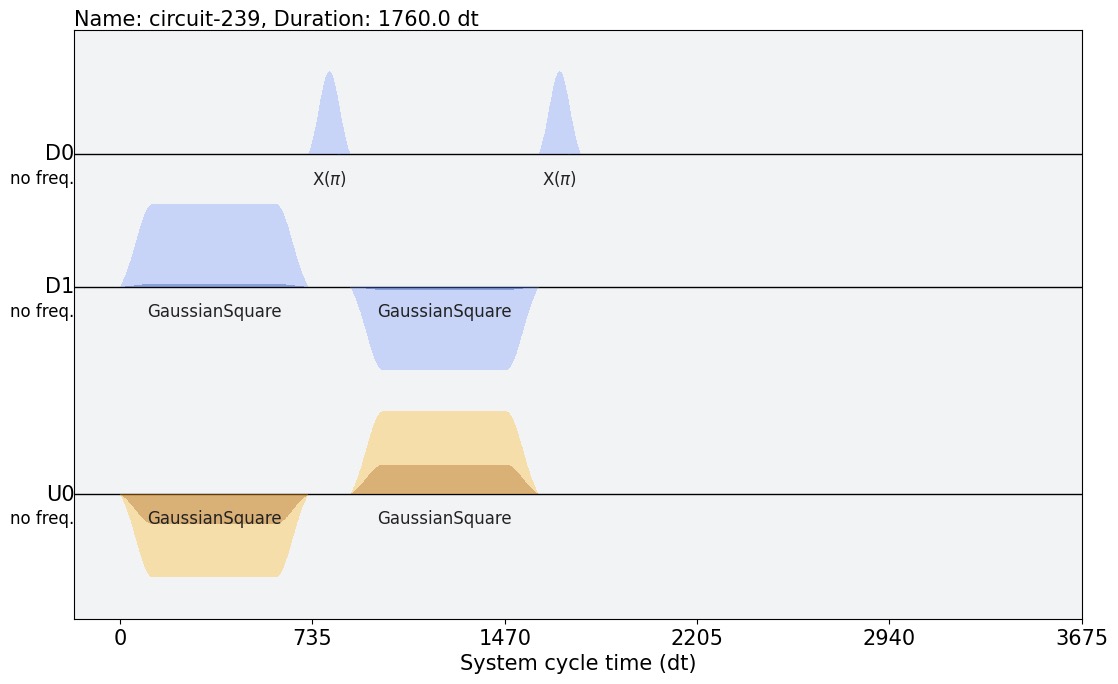}
    \vspace{-1em}
    \captionof{figure}{Pulses generated by SimuQ for evolving $Z_0Z_1$ for $T=1$.}
    \label{fig:pulse-simuq}
    \includegraphics[width=0.70\linewidth]{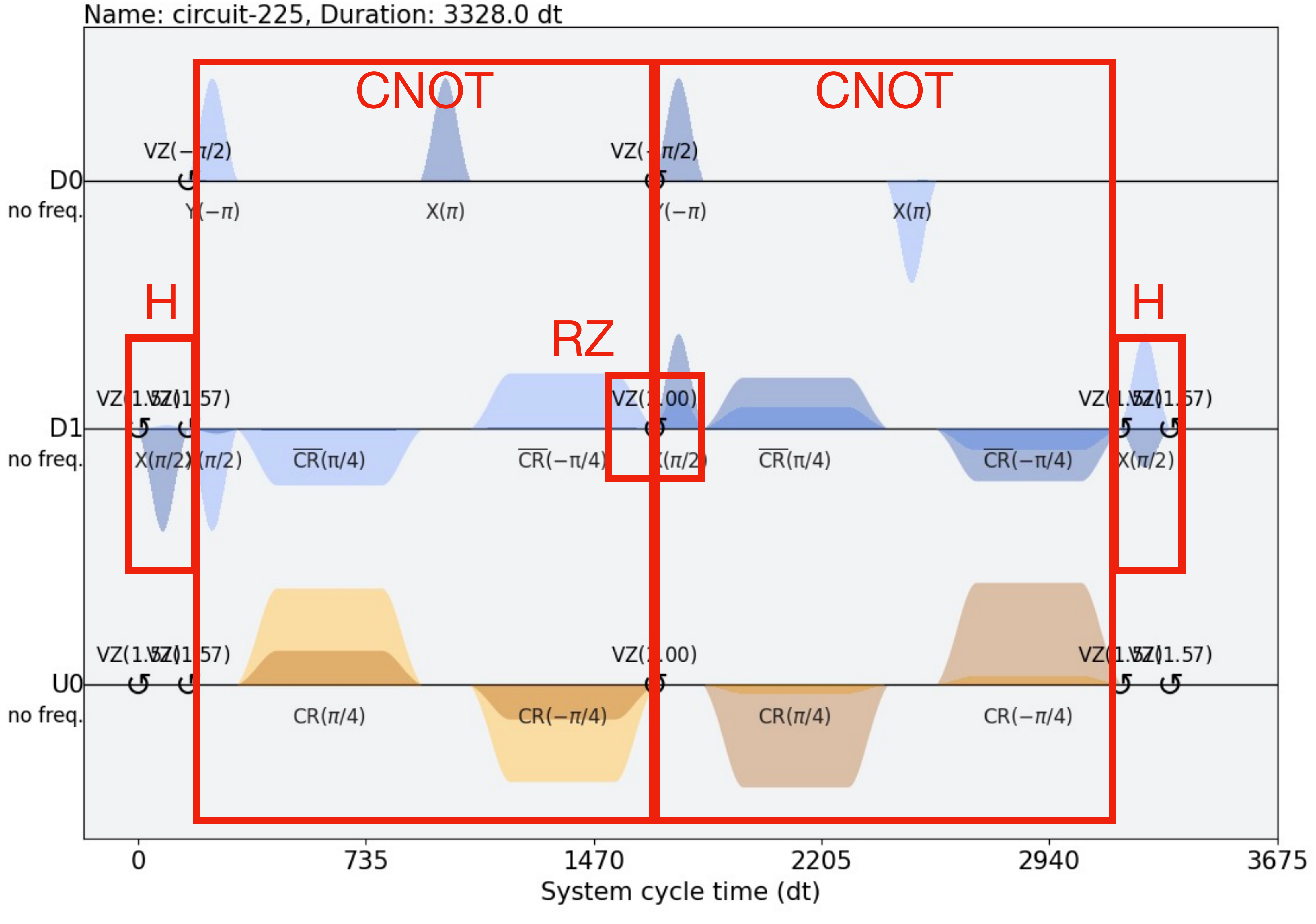}
    \vspace{-1em}
    \captionof{figure}{Pulses generated by Qiskit for evolving $Z_0Z_1$ for $T=1$.}
    \label{fig:pulse-qiskit}
    \vspace{1em}
\scalebox{0.9}{
\begin{tabular}{|c|cc|cc|}
\hline
\multirow{2}{*}{$p$} & \multicolumn{2}{c|}{IBM}                     & \multicolumn{2}{c|}{IonQ}                    \\ \cline{2-5} 
                   & \multicolumn{1}{c|}{SimuQ} & Qiskit & \multicolumn{1}{c|}{SimuQ} & Qiskit \\ \hline
1                  & \multicolumn{1}{c|}{\textbf{0.724}}      & 0.855      & \multicolumn{1}{c|}{0.240}      & \textbf{0.114}      \\ \hline
2                  & \multicolumn{1}{c|}{\textbf{1.365}}      & 1.929      & \multicolumn{1}{c|}{\textbf{
0.453}}      & 0.474      \\ \hline
3                  & \multicolumn{1}{c|}{\textbf{2.082}}      & 3.166      & \multicolumn{1}{c|}{\textbf{0.518}}      & 0.715      \\ \hline
\end{tabular}
}
\captionof{table}{The differences between the ideal $C(s)$ and the measured $C(s)$ on devices are displayed for different layers $p$ with SimuQ and Qiskit CNOT-based compiler and are better when lower. }
\label{tab:measured_cut}
\end{minipage}
}
\end{table}

%% file: tabs/benchmark.tex
\begin{table}
\scalebox{0.95}{
\begin{tabular}{|c|c|c|c|ccc|cc|}
\hline
\multirow{2}{*}{\textbf{\begin{tabular}[c]{@{}c@{}}System\\ name\end{tabular}}} & \multirow{2}{*}{\textbf{\begin{tabular}[c]{@{}c@{}}LoC\end{tabular}}} & \multirow{2}{*}{\textbf{\begin{tabular}[c]{@{}c@{}}\# of \\sites\end{tabular}}} & \textbf{QuEra}                                                    & \multicolumn{3}{c|}{\textbf{IBM}}                                                                                                                                                                                                                          & \multicolumn{2}{c|}{\textbf{IonQ}}                                                                                                                            \\ \cline{4-9} 
                                                                                &                                                                                  &                                                                                     & \textbf{\begin{tabular}[c]{@{}c@{}}Comp.\\ time $(s)$\end{tabular}} & \multicolumn{1}{c|}{\textbf{\begin{tabular}[c]{@{}c@{}}Comp.\\ time $(s)$\end{tabular}}} & \multicolumn{1}{c|}{\textbf{\begin{tabular}[c]{@{}c@{}}P.D. $(\mu s)$\\ SimuQ\end{tabular}}} & \textbf{\begin{tabular}[c]{@{}c@{}}P.D. $(\mu s)$\\ Qiskit\end{tabular}} & \multicolumn{1}{c|}{\textbf{\begin{tabular}[c]{@{}c@{}}Comp.\\ time $(s)$\end{tabular}}} & \textbf{\begin{tabular}[c]{@{}c@{}}\# of\\ 2q-gate\end{tabular}} \\ \hline
\multirow{4}{*}{ising\_chain}                                                   & \multirow{4}{*}{13}                                                              & 6                                                                                   & 0.177                                                             & \multicolumn{1}{c|}{0.224}                                                          & \multicolumn{1}{c|}{2.06}                                                                 & 8.69                                                                  & \multicolumn{1}{c|}{0.155}                                                             & 20                                                                   \\ \cline{3-9} 
                                                                                &                                                                                  & 32                                                                                  & 39.3                                                              & \multicolumn{1}{c|}{54.6}                                                              & \multicolumn{1}{c|}{3.24}                                                                 & 39.2                                                                  & \multicolumn{1}{c|}{47.2}                                                              & 124                                                                  \\ \cline{3-9} 
                                                                                &                                                                                  & 64                                                                                  & 663                                                               & \multicolumn{1}{c|}{257}                                                               & \multicolumn{1}{c|}{3.15}                                                                 & 81.2                                                                  & \multicolumn{1}{c|}{680}                                                               & 252                                                                  \\ \cline{3-9} 
                                                                                &                                                                                  & 96                                                                                  & 2298                                                              & \multicolumn{1}{c|}{1086}                                                              & \multicolumn{1}{c|}{3.26}                                                                 & 450                                                                   & \multicolumn{1}{c|}{3568}                                                              & 380                                                                  \\ \hline
\multirow{4}{*}{ising\_cycle}                                                   & \multirow{4}{*}{13}                                                              & 6                                                                                   & 0.585                                                             & \multicolumn{3}{c|}{No. sol.}                                                                                                                                                                                                                                  & \multicolumn{1}{c|}{0.13}                                                              & 24                                                                   \\ \cline{3-9} 
                                                                                &                                                                                  & 12                                                                                  & 3.47                                                              & \multicolumn{1}{c|}{1.49}                                                              & \multicolumn{1}{c|}{2.05}                                                                 & 37.8                                                                  & \multicolumn{1}{c|}{1.37}                                                              & 48                                                                   \\ \cline{3-9} 
                                                                                &                                                                                  & 32                                                                                  & 114                                                               & \multicolumn{1}{c|}{483}                                                               & \multicolumn{1}{c|}{3.35}                                                                 & 144                                                                   & \multicolumn{1}{c|}{53.8}                                                              & 128                                                                  \\ \cline{3-9} 
                                                                                &                                                                                  & 64                                                                                  & 3454                                                              & \multicolumn{3}{c|}{Time out}                                                                                                                                                                                                                              & \multicolumn{1}{c|}{907}                                                               & 256                                                                  \\ \hline
heis\_chain                                                                     & 15                                                                               & 32                                                                                  & No. sol.                                                              & \multicolumn{1}{c|}{143}                                                               & \multicolumn{1}{c|}{10.1}                                                                 & 119                                                                   & \multicolumn{1}{c|}{138}                                                               & 372                                                                  \\ \hline
qaoa\_cycle                                                                     & 19                                                                               & 12                                                                                  & No. sol.                                                              & \multicolumn{1}{c|}{0.503}                                                             & \multicolumn{1}{c|}{0.83}                                                                 & 37.6                                                                  & \multicolumn{1}{c|}{1.5}                                                               & 36                                                                   \\ \hline
qhd                                                                             & 16                                                                               & 16                                                                                  & No. sol.                                                              & \multicolumn{3}{c|}{No. sol.}                                                                                                                                                                                                                                  & \multicolumn{1}{c|}{66.3}                                                              & 480                                                                  \\ \hline
\multirow{2}{*}{mis\_chain}                                                     & \multirow{2}{*}{22}                                                              & 12                                                                                  & 5.45                                                              & \multicolumn{1}{c|}{19.1}                                                              & \multicolumn{1}{c|}{18.9}                                                                 & 94                                                                    & \multicolumn{1}{c|}{25.2}                                                              & 440                                                                  \\ \cline{3-9} 
                                                                                &                                                                                  & 24                                                                                  & 53.1                                                              & \multicolumn{1}{c|}{328}                                                               & \multicolumn{1}{c|}{18.9}                                                                 & 162                                                                   & \multicolumn{1}{c|}{278}                                                               & 920                                                                  \\ \hline
\multirow{2}{*}{mis\_grid}                                                      & \multirow{2}{*}{29}                                                              & 16                                                                                  & 28.4                                                              & \multicolumn{3}{c|}{No. sol.}                                                                                                                                                                                                                                  & \multicolumn{1}{c|}{85.4}                                                              & 960                                                                  \\ \cline{3-9} 
                                                                                &                                                                                  & 25                                                                                  & 141                                                               & \multicolumn{3}{c|}{No. sol.}                                                                                                                                                                                                                                  & \multicolumn{1}{c|}{489}                                                               & 1600                                                                 \\ \hline
kitaev                                                                          & 13                                                                               & 18                                                                                  & 4.67                                                              & \multicolumn{1}{c|}{15.6}                                                              & \multicolumn{1}{c|}{2.12}                                                                 & 21.2                                                                  & \multicolumn{1}{c|}{8.74}                                                              & 68                                                                   \\ \hline
schwinger                                                                       & 18                                                                               & 10                                                                                  & No. sol.                                                              & \multicolumn{3}{c|}{No. sol.}                                                                                                                                                                                                                                  & \multicolumn{1}{c|}{1.09}                                                              & 28                                                                   \\ \hline
o3nl$\sigma$m                                                                          & 19                                                                               & 30                                                                                  & No. sol.                                                              & \multicolumn{3}{c|}{No. sol.}                                                                                                                                                                                                                                  & \multicolumn{1}{c|}{77.7}                                                              & 588                                                                  \\ \hline
\end{tabular}
}
\caption{A benchmark of quantum simulation problems. We program and compile the models in SimuQ to obtain pulse schedules for QuEra and IBM devices and quantum circuits for IonQ devices. We record the compilation time (comp. time), the pulse duration (P.D.), and the 2-qubit gate count for the generated circuits. No. sol. represents cases where the SimuQ compiler reports no solution because of hardware constraints, such as limited interactions of QuEra devices and machine topology for IBM devices. Time out is reported when the compilation takes more than an hour, which happens in the search for a 64-qubit cycle on IBM devices. }
\vspace{-2em}
\label{tab:benchmark}
\end{table}

%% file: 6-Conclusion.tex
\section{Conclusion and Future Directions}
\label{sec:future-work}

The domain-specific language SimuQ described in this paper is the first framework to consider quantum simulation and compilation to multiple platforms of analog quantum simulators. We propose HML for front-end users to program their target quantum systems intuitively. We also design abstract analog instruction sets to depict the programmability of analog quantum simulators and the AAIS-SL to program them. Furthermore, the SimuQ compiler is the first compiler to generate pulse schedules of analog quantum simulators for desired quantum simulation. 

Since this is the first feasibility demonstration of programming analog quantum simulators, there is much optimization space for our compiler. First, since different devices have different properties crucial to the compiler's efficiency, we can develop compilation passes specifically for each platform. Second, this paper employs a brute-force search with heuristics to find a site layout where more pruning techniques are desired. Third, The hand-crafted mixed-binary equation solver can also be optimized according to the structure of the problem. Furthermore, with a better understanding of hardware, we can design more powerful AAISs. Lastly, we can add more compilation techniques like \cite{clinton2021hamiltonian} to synthesize product Haimltonians not appearing directly in the given AAIS with a combination of instruction executions.

%% file: 9-Appendix.tex
\section{Details of the Equation System in Running Example}
\label{app:eqset}
\begin{align}
        \frac{C}{4(x_1-x_2)^6}&=1 \tag{$Z_1Z_2$}\\
        \frac{C}{4(x_2-x_3)^6}&=1 \tag{$Z_2Z_3$}\\
        \frac{C}{4(x_1-x_3)^6}&=0 \tag{$Z_1Z_3$}\\
        -\frac{C}{4(x_1-x_2)^6}-\frac{C}{4(x_1-x_3)^6}+\frac{\Delta_{\iota_1}}{2}\cdot s_{\iota_1}&=0 \tag{$Z_1$}\\
        -\frac{C}{4(x_1-x_2)^6}-\frac{C}{4(x_2-x_3)^6}+\frac{\Delta_{\iota_2}}{2}\cdot s_{\iota_2}&=0 \tag{$Z_2$}\\
        -\frac{C}{4(x_1-x_3)^6}-\frac{C}{4(x_2-x_3)^6}+\frac{\Delta_{\iota_3}}{2}\cdot s_{\iota_3}&=0 \tag{$Z_3$}\\
        \frac{\Omega_{\iota_1}\cos(\phi_{\iota_1})}{2}\cdot s_{\iota_1}&=1 \tag{$X_1$}\\
        \frac{\Omega_{\iota_2}\cos(\phi_{\iota_2})}{2}\cdot s_{\iota_2}&=1 \tag{$X_2$}\\
        \frac{\Omega_{\iota_3}\cos(\phi_{\iota_3})}{2}\cdot s_{\iota_3}&=1 \tag{$X_3$}\\
        \frac{\Omega_{\iota_1}\sin(\phi_{\iota_1})}{2}\cdot s_{\iota_1}&=0 \tag{$Y_1$}\\
        \frac{\Omega_{\iota_2}\sin(\phi_{\iota_2})}{2}\cdot s_{\iota_2}&=0 \tag{$Y_2$}\\
        \frac{\Omega_{\iota_3}\sin(\phi_{\iota_3})}{2}\cdot s_{\iota_3}&=0 \tag{$Y_3$}
\end{align}

\section{Optimization in Trotterization}
\label{app:trotter-opt}

Although Trotterization provides a theoretical guarantee of dealing with conflicts, in practice, to reduce the approximation errors, we need a large Trotterization number which also amplifies device noises. Many optimization techniques may be used in Trotterization, and we implement the following two as a demonstration. We leave further optimizations as future directions.

\paragraph{Order inside a Trotterization step} Notice that the error bound in \cite{pnas-simulation} does not require a fixed simulation order of $H_j$. We realize the freedom here by setting the blocks in one Trotterization step to be parallel in the temporal graph and connecting the blocks in the next step after the blocks in this step. This freedom may be exploited in the following compilation passes.

\paragraph{Blocks commuting with others} If the Hamiltonian of a block commutes with other blocks, we separate it from others, evolve it first, and resolve the conflicts of others. This procedure does not introduce errors.

\section{Benchmark Details}
\label{app:benchmark}

We present more details here for our benchmark. Each quantum system consists of a Hamiltonian, described below, and an evolution time $t\in[0, T)$ where $T=1$. Whenever Trotterization is needed, we set the Trotterization number $R=4$ because of our empirical observations that this is a balance between Trotterization error and device noises. For the time-dependent Hamiltonian, we set the discretization number $D=10$.

Our benchmark includes:
\begin{itemize}
    \item \textbf{ising\_chain}: This is a transverse-field chain Ising model with $n$ sites
    \begin{align}
        H(t)=J\sum_{j=1}^{n-1}Z_jZ_{j+1} + h\sum_{j=1}^n X_j.
    \end{align}
    
    \item \textbf{ising\_cycle}: This is a transverse-field cycle Ising model with $n$ sites
    \begin{align}
        H(t)=J\left(Z_1Z_n+\sum_{j=1}^{n-1}Z_jZ_{j+1}\right) + h\sum_{j=1}^n X_j.
    \end{align}
    
    \item \textbf{heis\_chain}: This is a chain Heisenberg model with $n$ sites
    \begin{align}
        H(t)=J\sum_{j=1}^{n-1}\left(X_jX_{j+1}+Y_jY_{j+1}+Z_jZ_{j+1}\right) + h\sum_{j=1}^n X_j.
    \end{align}
    
    \item \textbf{qaoa\_cycle} \cite{farhi2014quantum}: This is an algorithm finding the Max-Cut of a graph $E$. It is an alternative evolution of $p$ layers under $H_1$ and $H_2$ with $n$ sites
    \begin{align}
        H_1(t)=\sum_{(j, k)\in E}Z_jZ_k, \quad H_2(t)=\sum_{j=1}^n X_j.
    \end{align}
    We set $E$ to be a $n$-site cycle.
    Note that in the algorithm the initial state is $\ket{+}^{\otimes n}$, which can be realized by evolving under $H_0$ for $\pi/\sqrt{8}$ where
    \begin{align}
        H_0(t)=\sum_{j=1}^n (X_j+Z_j).
    \end{align}
    The measurement results give a partition of the vertices in the graph approximating the maximal cut.
    
    \item \textbf{qhd} \cite{leng2023quantum}: This is an algorithm finding the minimal point of $f(x, y)$ in $[0, 1]^2$, which is an evolution under
    \begin{align}
        H(t)=-\frac{1}{2} \alpha \nabla^2 + \beta f(\hat{x}, \hat{y}).
    \end{align}
    We set $q=8, f=-x^2+y^2+xy$, and use two $q$-bit quantum register to represent $x$ and $y$, whose qubits are labeled by $(x, j)$ and $(y, j)$ correspondingly.
    Then
    \begin{align}
        \nabla^2&=q^2 \sum_{j=1}^q (X_{x,j} + X_{y,j}), \\
        \hat{x}&=\frac{1}{q} \sum_{j=1}^q n_{x,j}, \\
        \hat{y}&=\frac{1}{q} \sum_{j=1}^q n_{y,j}.
    \end{align}
    
    \item \textbf{mis\_chain} \cite{ebadi2021quantum}: This algorithm finds the maximal independent set of a graph $E$. It is an evolution under
\begin{align}
    H(t)=\sum_{i=1}^{N} (-\delta(t)\hat{n}_i+\frac{\omega}{2}X_i)+\sum_{(i, j)\in E}\alpha\hat{n}_i \hat{n}_j
\end{align}
for $N=|V|$, time interval $[0, 1],$ and $\delta(t)=(-1+2t)U$. Here we set $E$ an $N$-vertex chain, and $U, \omega,$ and $\alpha$ are real amplitude constants designed in the encoding \cite{ebadi2022quantum}. The quantum state's measurement result at the evolution's end encodes an approximate maximum independent set of the graph (after post-processing).

    \item \textbf{mis\_grid} \cite{ebadi2021quantum}: This is similar to the above one. The graph $E$ is a grid graph where the vertices are labeled by $(i, j)$ and there is an edge between $(i, j)$ and $(i+1, j)$, and an edge between $(i, j)$ and $(i, j+1)$.

    \item \textbf{kitaev} \cite{aramthottil2022scar}: This is a model to research weak ergodicity breaking induced by quantum many-body scars. It is an evolution of $n$ sites under
    \begin{align}
        H = \frac{\mu}{2} \sum_{j=1}^{n-1} Z_jZ_{j+1} - \sum_{j=1}^n (t X_j + h Z_j)
    \end{align}

    \item \textbf{schwinger} \cite{hamer1997series}: This is a model for lattice gauge theory in high-energy physics. It is an evolution of $n$ sites under:
    \begin{align}
        H = \frac{m}{2} \sum_j (-1)^j Z_n + \frac{\omega}{2} \sum_{j=1}^{N-1} (X_jX_{j+1}+Y_jY_{j+1}) + J \sum_{j=1}^{N-1} (\epsilon_0 + 1/2 \sum_{k=1}^j (Z_k+(-1)^k))^2
    \end{align}

    \item \textbf{o3nl$\sigma$m} \cite{ciavarella2023preparation}: This is a simulation of (1+1D) $O(3)$ non-linear $\sigma$ model. It is an evolution of $n\times m$ sites under:
    \begin{align}
        H = J_x \sum_{(j, k)} \sigma_{j, k}\cdot\sigma_{(j+1), k} + J_y \sum_{(j, k)} \sigma_{j, k}\cdot\sigma_{j, (k+1)}.
    \end{align}
    Here the sites are labeled by $(j, k)$, and $\sigma_{j, k}=(X_{j, k}, Y_{j, k}, Z_{j, k})$.
    
\end{itemize}